 \documentclass[final,authoryear,5p]{elsarticle}

\usepackage{epsfig}

\usepackage{amssymb}

\usepackage[ps2pdf,%
a4paper=true,%
breaklinks=true,%
colorlinks=true,%
pdfauthor={Radhika \& Nandi},%
pdftitle={Template for manuscripts in Advances in Space Research}%
]{hyperref}

\journal{Advances in Space Research}

\begin{document}

\begin{frontmatter}



\title{`Spectro-temporal' characteristics and disk-jet connection of the outbursting black hole source 
XTE J1859$+$226} 


\author{Radhika. D$^{1,2}$\corref{cor}}
\address{$^1$ Space Astronomy Group, SSIF, ISRO Satellite Centre, Bengaluru, INDIA \\ $^2$ Department 
of Physics, University of Calicut, Kerala, INDIA}
\cortext[cor]{Corresponding author}
\ead{radhikad\_isac@yahoo.in}


\author{A. Nandi$^1$}
\address{$^1$Space Astronomy Group, SSIF, ISRO Satellite Centre, Bengaluru, INDIA}
\ead{anuj@isac.gov.in}

\begin{abstract}
We re-investigated the `spectro-temporal' behaviour of the source XTE J1859$+$226 in X-rays during 
its outburst phase in 1999, by analysing the RXTE PCA/HEXTE data in 2 - 150 keV spectral band. 
Detailed analysis shows that the source evolves through different spectral states during its entire 
outburst as indicated by the variation in the spectral and temporal characteristics. Although the 
evolution pattern of the outburst followed the typical q-shaped profile, we observed an absence of 
`canonical' soft state and a weak presence of `secondary' emission during the decay phase of the 
outburst. The broad-band spectra, modeled with high energy cutoff, shows that the fold-energy increases 
monotonically in the hard and hard-intermediate states followed by a random variation in the 
soft-intermediate state. We attempted to estimate the mass of the source based on the evolution of 
Quasi-Periodic Oscillation (QPO) frequencies during rising phase modeled with the propagating 
oscillatory shock solution, and from the correlation of photon index and QPO frequency. It is also 
observed that during multiple ejections (observed as radio flares)
the QPO frequencies are not present in the power spectra and there is an absence of lag in the soft 
to hard photons. The disk flux increases along with a decrease in the high energy flux, implying the 
soft nature of the spectrum. 
These results are the `possible' indication that the inner part of the disk ({\it i.e., 
Comptonized corona}), which could be responsible for the 
generation of QPO and for the non-thermal Comptonized component of the spectrum, is disrupted and 
the matter gets evacuated in the form of jet. We attempted to explain the complex behaviour 
of `spectro-temporal' properties of the source during the entire outburst and the nature of the 
disk-jet connection before, during and after the ejection events in the context of two 
different types of accreting flow material, in presence of magnetic field.

\end{abstract}


\end{frontmatter}

\parindent=0.5 cm

\section{Introduction}

Galactic Black Hole (BH) sources are interesting objects to study as these sources are observed only 
in binaries and the process of accretion gets very complex as the disk evolves with time, especially 
when the sources undergo outbursts and jet ejections take place.
Some of the BH binaries show persistent emission (e.g. Cyg X$-$1) along with aperiodic X-ray
variability (e.g. GRS 1915$+$105), over more than a period of decade. Some BH sources show 
outbursting behaviour for shorter duration varying from few days to months, and are called as 
outbursting BH sources or transients (e.g. GX 339$-$4, XTE J1118$+$480, XTE J1748$-$288, H 1742$-$322, 
GRO J1655$-$40). The outbursting BH sources show different types of intensity variations 
(i.e., outburst profile) over different period of time \citep{RM04}. Some of the sources after being 
quiescent for a long time, show a sudden increase in the intensity level, and attain a maximum 
intensity within few days and then decay back slowly to quiescence (H 1743$-$322, A 0620$-$00, 
4U 1543$-$47). Their light curves have a `Fast Rise and Exponential Decay (FRED)' 
profile. Some sources have a slow rise to the peak and decay slowly to the quiescence 
(GX 339$-$4, XTE J1752$-$223) and their light curve profile is termed as `Slow Rise Slow 
Decay (SRSD)'. During this whole phenomena, outbursting sources show different spectral and timing 
variabilities and exhibit different spectral states like {\it Hard, Hard-intermediate, 
Soft-intermediate, Soft state} (\citealt{HB2005}, \citealt{TB2010}) and in some cases 
a {\it Very high state} also (\citealt{Miyamoto91}, \citealt{Rem1999}), in their 
Hardness-Intensity diagram (HID).

It is also observed that in outbursting BH sources, strong jets are emitted which are seen in radio 
observations \citep{FBG04}, during the transition from hard-intermediate to 
soft-intermediate state in the rising phase (e.g. H 1743$-$322, \citealt{h2012}) and in the declining 
phase when the source transits from soft-intermediate to hard state (e.g. XTE J1752$-$223, 
\citealt{Yang2010}).
Quasi-simultaneous multiwavelength observations of sources like GX 339$-$4 \citep{CB2011}, 
XTE J1748$-$288 \citep{Brocksopp2007}, H 1743$-$322 \citep{h2012}, strongly suggests that the radio 
flares emitted are associated with the disk emission. It has been reported from the study of 
GRS 1915$+$105 that, during the jet ejections, QPOs are not observed as well
as the Comptonized component gets suppressed (\citealt{SV2001}). This implies that the `hot' 
Comptonized corona gets disrupted and evacuated, and the source spectra 
softens, suggesting that the X-ray emission is mostly from the disk (\citealt{Feroci99}, 
\citealt{SV2001}, \citealt{Nandi2001}, \citealt{Chak2002}, \citealt{h2012}). 

The X-ray transient source XTE J1859$+$226 was first discovered \citep{Wood99} with All Sky Monitor 
(ASM) onboard Rossi X-ray Timing Explorer(RXTE) \citep{BRS93} on Oct 9, 1999. Subsequently, the source 
was monitored in X-rays with RXTE/PCA (Proportional \\Counter Array) and CGRO-BATSE (Compton Gamma Ray 
Observatory - Burst And Transient Source Experiment) for several months \citep{McW99}. 
The outburst showed the typical FRED profile. 
Spectral and temporal characteristics confirmed the source as a black hole candidate 
\citep{Mark2001}. Several observations in optical and radio wavebands confirmed the presence of 
the counterpart of the source \citep{Garn99, Pooley99}. Spectroscopic studies of the counterpart 
showed weak emission lines arising from Balmer series of Hydrogen and He II, which is typical for 
spectra of LMXBs \citep{Wagner99}. The mass function was estimated to be (4.5 +/- 0.6) 
M$_{\odot}$, and an assumed inclination angle of 70$^\circ$ gave a lower mass 
limit of 5.42 M$_\odot$ \citep{Corral2011}.

During the 1999 outburst, the source XTE J1859$+$226 was continuously and extensively monitored in 
X-rays and in radio, which revealed the X-ray/radio correlations \citep{Brocksopp2002}. 
The source is observed to exhibit multiple flaring events of five in number. 
From the study of the spectral evolution in radio, it was found that the jet generation 
is implied by the production of hard X-rays and also that a correlation exists between soft X-ray 
and radio ejection during the first flare, while a correlation existed between hard X-ray and 
radio observation during the other ejections.
\citealt{Casella2004} studied the temporal properties of the source and found that, the QPOs 
observed can be classified into three types viz. Type A, B and C. This classification scheme based on 
the QPO characteristics and phase lag, has been considered as one of the basic formalities to 
classify the QPOs in Black Hole sources (See also \citealt{Cas2005, Motta2011}). 
The phase lag difference between different types of QPOs, suggests that the shape of oscillation is 
different in different energies. But the evolution of low frequency QPOs (C-type) as well as 
their origin during the initial rising phase of the outburst is still not clear. 

In the context of disk-jet symbiosis of black hole sources, \citealt{FBG04,FHB09} 
have provided a unified picture along with the estimation of jet power as a function of X-ray 
luminosity.
Their work also highlighted the occurrence of five Radio flares in XTE J1859$+$226
and suggested that all the flares occur when the source is in a similar spectral state. They
found that the rms of the Power density spectra (PDS) reduces during the occurrence of a flare.
\citealt{Rodriguez2008} studied the spectral properties of the source during its rising phase 
and found that the non-thermal flux (2 - 50 keV) remains constant during the first flare of the 
source. \citealt{Mark2001} studied the evolution of the source during the rising phase of the 
outburst and found that there is a signature of partial absence of the 
QPO (or Comptonized corona), when the first radio flare occurred.

\citealt{Dunn2011a,Dunn2011b} studied the spectral behaviour of this source in the context 
of global study of disk dominated states of several BH sources. Considering a simple phenomenological 
model (diskbb+powerlaw) for BH spectral study, \citealt{Dunn2011a} showed that a large fraction of 
disk dominated observations of XTE J1859$+$226 fall below the `standard' Luminosity-Temperature 
(L-T) relation (see \citealt{GD2004}). \citealt{Fari2013} has studied the spectral 
characteristics of the source based on BeppoSAX and RXTE observations, in the context of a bulk motion 
comptonization model (BMC of XSPEC), and have observed an evolving high energy cut-off 
component during the hard to soft transition. They have also performed a correlated study between the 
fraction of Comptonization and the rms variability. But the `complex' evolution of fold energy since 
the beginning of the outburst along with the evolution of 
temporal features (QPOs, rms etc.), especially during a radio flare are not explained in detail.

So, in order to have a coherent study of the evolution of temporal and spectral 
properties of the source (during the entire outburst) as well as the implications on disk dynamics 
during the multiple ejections, we re-analysed the RXTE PCA/HEXTE temporal and spectral data of the 
source XTE J1859$+$226 in the energy band of 2 - 150 keV. 

Several models have been proposed to understand the `spectro-temporal' evolution of an 
outbursting black hole binary. The accretion-ejection instability model \citep{TagPel} and global disk 
oscillation model \citep{TitOsh2000} attempts to explain the origin of QPOs.
Recently, \citealt{ID2011} based on propagating mass accretion rate fluctuations in hotter inner 
disk, and \citealt{Stiele2013} based on oscillations from a transition layer in between the disk and 
hot Comptonized flow have proposed alternative models to explain the origin of QPOs. 
There have been other attempts also to understand the hard X-ray spectral state variations 
\citep{Esin1997, Titarchuk2007, Motta2009, Dunn2011a, Dunn2011b} as well as 
state transitions across the HID \citep{Meyer2007,Meyer2009} of the outbursting 
BH sources.
But none of these models, addressed as a whole the issues of evolution of low frequency QPOs as well
as soft and hard X-ray spectral components, and spectral state changes across the 
HID during the entire outburst. On the other hand, quite independently, several alternative models
(see \S 4 for discussion) and more recently the phenomenological model \citep{FBG04,FHB09} have been put 
forward only for understanding the disk-jet symbiosis in BH sources.

So, we attempted to understand the evolution of temporal and spectral properties associated with the 
different branches of the `q-diagram' of the outburst of the source XTE J1859$+$226 within
a single framework of the Two Component Advective Flow (TCAF) model \citep{ST95}. 
Basically, this model consists of two different types of accreting flow 
\citep{ST95}: sub-Keplerian (i.e., freely falling and less viscous flow) and Keplerian 
(i.e., moves in circular orbit and high viscous flow) matter \citep{SS73}.
We also delve deep into the nature of the accretion dynamics during the Jet ejections of this
source based on the detailed temporal and spectral X-ray properties.

In general, outbursting black hole sources show signature of evolution of QPO frequencies 
\citep{BH1990, TB2005, Deb2008, Nandi2012}, 
which could be explained based on the Propagating oscillatory shocks 
solution \citep{skc08,skc09, Nandi2012} of TCAF model. 
In this work, we provided the solution of evolution of QPOs during the rising phase and 
possible explanation of the `reverse' nature of the evolution of the 
hard spectral component (i.e., the fold energy), which can be interpreted with the
shock acceleration mechanism \citep{ChakSam2006} in the sub-Keplerian flow.
Our findings show that the QPOs are not observed during the radio flares, which we have explained with 
the possible scenario of `evacuation/disruption' of the Comptonized corona. We attempt to 
estimate the mass of the source by modeling the QPO evolution and also from the correlation between 
QPO frequency and photon index \citep{Shap2007}. The preliminary results of our findings were 
presented in \citealt{NR2012}. 

This paper has been organized in the following manner: In the next section, we discuss the 
observations and the procedures applied for data analysis using the standard packages for RXTE 
PCA/HEXTE. In \S 3, we presented the results obtained from the temporal and spectral analysis for 
the whole outburst, estimation of mass in \S 3.2, followed by specific spectral and temporal 
characteristics during each radio flare in \S 3.3.
In \S 4, we discuss the possible physical scenario to explain the evolution of X-ray properties 
observed during the entire outburst of XTE J1859$+$226 and the disk-jet connection, and summarize
our conclusions.

\section {Observation and Data Analysis}

We analysed the public archival data obtained from the HEASARC database for the RXTE satellite to 
study the characteristics of the source XTE J1859$+$226 over the entire duration of the single outburst 
which occurred during the entire RXTE era. We analysed the PCA (2 - 25 keV) and HEXTE (20 - 150 keV) 
data spanning 166 days since October 9, 1999 (MJD 51460.76) to March 23, 2000 (MJD 51626.6). We 
excluded the observations of July 2000 when the source was in quiescence, as the counts were very 
less. The standard FTOOLS package of HEASOFT v 6.11 was used for data reduction. For spectral 
and temporal analysis purposes, we used the packages XSPEC v 12.7 and XRONOS v 5.22 respectively.
 
For the timing analysis, we used the Science data of PCA in the Binned mode (B\_8ms\_16A\_0\_35\_H\_4P,
FS37*) of maximum time resolution of 8 ms which spans over 0 - 35 channels, and Event mode 
(E\_16us\_16B\_36\_1s, FS3b*) of 16 $\mu$s time resolution for 36 - 249 channels. 
We used the FTOOLS task {\it xtefilt} to create filter file, and {\it maketime}
to generate a good time interval ({\it gti}) file, with conditions of {\it elevation angle $>$ 
10$^{\circ}$, offset of $<$ 0.02 and time since SAA passage $<$ 30 min}. Since it was not possible 
to select the data only for a single PCU from the Binned mode data, we decided to extract light 
curves for all PCUs. Light curves were generated in various energy bands of 2 - 6 keV, 
6 - 13 keV, 13 - 25 keV and 2 - 25 keV in order to perform an energy dependent study of the 
PDS. Since the Binned mode data consisted of only 0 - 35 channels (2 - 13 keV), 
we generated the light curve of minimum binning time, for the channel 
ranges 0 - 15 (2 - 6 keV) and 15 - 35 (6 - 13 keV), using {\it saextrct}. The light curve from the 
Event mode data was created using {\it seextrct} for the minimum available binning time, and the 
data belonging to channel range of 36 - 67 (13 - 25 keV) was extracted from the total available 
channel range of 36 - 249. A combined light curve for 2 - 25 keV was obtained, by summing up the 
light curves from 0 - 35 channels (2 - 13 keV) and 35 - 67 channels (13 - 25 keV), using 
{\it lcmath}. Since the contribution of background counts were very less (8.6 counts over 
2674 source counts at the peak of the outburst and 6.6 
counts over 61.9 source counts at the minimum), we did not subtract the background counts 
while generating the light curve.
 
For the spectral analysis, we decided to extract data for PCU2 only, since throughout 
the entire outburst only PCU2 remained `ON' and the efficiency of the detector also holds good.
The source spectra were extracted from the Standard2 data product (FS4a*), which has a time 
resolution of 16 sec. Background spectra were obtained with the help of {\it runpcabackest}, applying 
the background model for bright source and SAA passage history file obtained from the PCA background 
web-page\footnote{http://heasarc.gsfc.nasa.gov/docs/xte/pca\_bkg\_epoch.html} of RXTE for the 
corresponding epoch during which the source has undergone outburst. Spectral response was created 
using the {\it pcarsp} tool. We also generated the 
background subtracted light curve for PCU2 counts for the energy range 2 - 6 keV, 6 - 20 keV 
and 2 - 20 keV, in order to plot the HID.

We extracted high energy (20 - 150 keV) spectral data mostly from cluster A (FH52*) of HEXTE, 
since specific observation by cluster B (FH58*) had lesser counts, except for initial two 
days (MJD 51462.76 and MJD 51463.83), where there were no cluster A observation. Using standard 
procedures, we generated the deadtime corrected source and background spectra, and spectral response 
was generated using {\it hxtrsp}.

\begin{figure}
\hspace{-0.6cm}
\includegraphics[height=8cm,width=10cm]{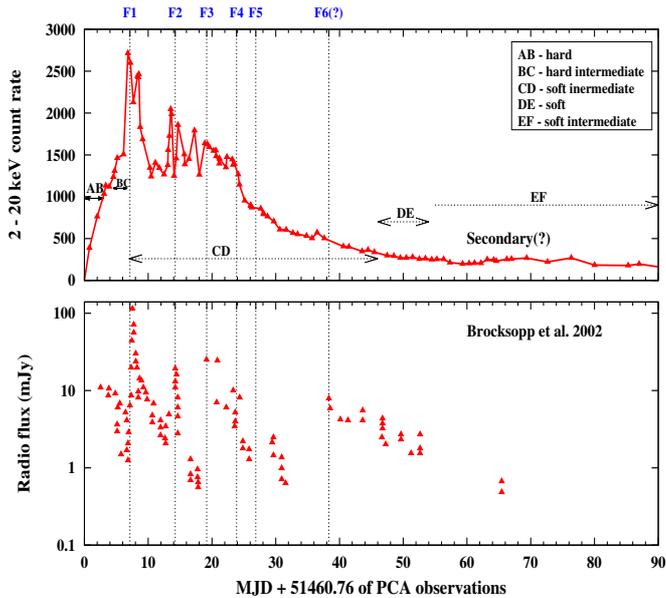}
\caption{2 - 20 keV light curve from PCU2 data is shown in the top panel followed by the radio 
light curve \citep{Brocksopp2002} in the bottom panel. 
We also mark the five flares \emph{peak} 
as F1, F2, F3, F4, F5. Please see text for flares F5 and F6(?) as there were no radio 
observation during F5 and before F6. The different spectral states (except the hard intermediate (FG) 
and hard states (GH) in declining phase) are also shown.}
\label{rlc-pca}
\end{figure}

Radio observations were carried out in the bands of 1.43 GHz, 1.66 GHz, 2.25 GHz, 3.9 GHz by 
various observatories (VLA, MERLIN, GBI, RATAN) during the outburst of the source, and the 
combined radio light curve was presented in \citealt{Brocksopp2002}. We re-produced the radio 
light curve presented in \citealt{Brocksopp2002} using the software `DEXTER' 
\footnote{http://dexter.edpsciences.org/Dexterhelp.html}. Figure \ref{rlc-pca} shows the X-ray and 
the radio light curve for the source XTE J1859$+$226 during the entire outburst. 

\citealt{Brocksopp2002} (bottom panel of Figure \ref{rlc-pca}) identified five radio flares during 
the entire outburst. The five flares were observed to have their {\it peak} radio flux on 
MJD 51467.90 (F1), 51475.00 (F2), 51479.94 (F3), 51484.62 (F4), 51487.63 (F5).  

\subsection{Timing analysis}

To obtain the PDS from the lightcurves generated, we used 
the `powspec v 1.0' of the XRONOS package and the customized 
software based on IDL, {\it General High energy Aperiodic Timing Software} (GHATS) v 1.0.1 
\footnote{http://www.brera.inaf.it/utenti/belloni/GHATS\_Package/Home.html} . While using `powspec', 
a normalisation factor 
of `-2' was applied in order to subtract the expected white noise level from the data, to obtain the 
squared rms fractional variability from the integral of the PDS. Further analysis shows that the 
factor of `-2' deviates about 0.5\% in almost all states except the soft-intermediate states where 
it deviates by 3\% to 5\%. The GHATS package considers dead time effect and hence the correct value of 
normalisation factor (as in \citealt{Zhang1995}) while generating the PDS. The available minimum time 
binning factor of 0.0078 sec, which corresponds to a Nyquist frequency of 64 Hz, was chosen over 
8192 segments. The power obtained in the PDS has units of rms$^2$/Hz. 

We modeled the power spectra from 0.1 to 64 Hz, by different components of 
{\it Lorentzians} (\citealt{Belloni2002}) for the QPOs and broad-band noise, and a powerlaw 
for the red noise, wherever required. The resultant centroid frequency is considered as 
the QPO frequency. 
We also estimated the Q-factor (centroid frequency/width of the Lorentzian), amplitude in rms and 
significance of the QPO observed. The standard QDP/PLT command 
of {\it statistic} was used to obtain the integrated value (rms) for the QPO and for the overall PDS, 
 in the range of 0.1 to 64 Hz. 
The error values on each parameter of the components were estimated at 90\% confidence interval, 
using the {\it fit err} command.

We also studied the evolution of PDS in different branches of HID of the outburst. The phase 
lag between soft (2 - 6 keV) and hard (6 - 25 keV) photons during the flares was estimated using 
GHATS. The lightcurves and hence the Fast Fourier Transforms for the 2 - 6 keV and 6 - 25 keV energy 
bands are generated. The cross spectrum is computed, taking into account the subtraction of Poissonian 
noise (as per \citealt{Zhang1995}). We followed the standard procedures of the GHATS package to 
estimate the lag (see also \citealt{Casella2004} \& Figure \ref{lag-F3} of this work).

\subsection{Spectral analysis} 

We performed a simultaneous fit of the PCA and HEXTE data for the energy range of 3 - 150 keV, 
including a constant factor close to unity for normalising both the spectra. In order to get an idea of 
the normalising factor we estimated the factor by simultaneously fitting the Crab spectra (since it 
does not vary with time) observed by RXTE during the different spectral states of XTE J1859$+$226 
(observation IDs 40093-01-12-00 and 40093-01-15-00). The constant was observed to be $\sim$ 1. A 
simultaneous fit to the PCA and HEXTE spectra of XTE J1859$+$226 showed that when the constant factor 
was fixed at 1 for PCA, a value varying between 0.9 and 1.1 (over the entire data set) was obtained 
for the constant factor corresponding to HEXTE spectra, which was used during the fitting.

We modeled the energy spectra using the standard accretion disk models ({\it diskbb} and 
{\it powerlaw}) for BH spectra, modified by the \textit{phabs} model to account for the interstellar 
absorption. 
From an initial fit to the data sets during the entire outburst, we found that the nH parameter has 
an average value of 0.2 $\times$ 10$^{22}$ atoms cm$^{-2}$ and fixed this value for further analysis. 
This value of nH, agrees well with the average value of 0.216 $\times$ 10$^{22}$ 
atoms cm$^{-2}$, estimated by the Leiden/Argentine/Bonn (LAB) survey, and the Dickey \& Lockman (DL) 
survey value of 0.221 $\times$ 10$^{22}$ atoms 
cm$^{-2}$(\footnote{http://heasarc.nasa.gov/cgi-bin/Tools/w3nh/w3nh.pl}). \citealt{Mark2001} 
quoted a slightly higher value of 1.1 $\times$ 10$^{22}$ atoms cm$^{-2}$. We find that a change in 
nH from 0.2 to 1.1, does not significantly change the overall results. 

\begin{figure}
\includegraphics[height=8cm,width=5cm,angle=-90]{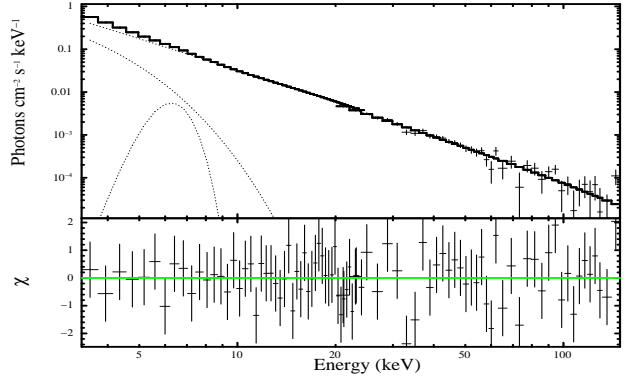}
\caption{Unfolded PCA (3 - 20 keV) and HEXTE (20 - 150 keV) spectra of the X-ray observation of 24 hrs 
before the occurrence of first flare (MJD 51466.89, observation ID: 40124-01-11-00). The combined 
spectra of 3 - 150 keV is modeled by a diskbb ($T_{in}=0.89$ keV), powerlaw ($\Gamma=2.35$), a high 
energy cutoff ($E_{fold}=151.28$ keV) along with Gaussian \& smedge components.}
\label{ufspec-highecut}
\end{figure}

For an initial fit to the observation 40124-01-11-00 with the model {\it phabs*(diskbb+powerlaw)const}, 
we obtained $\chi^{2}_{red}$ (=$\chi^{2}$/dof) of 145.43/82. This resulted in residuals around 6.4 keV 
along with a smeared edge of $\sim$ 8 keV \citep{Ebisawa1994} and the 
model considered was \\ 
\textit{phabs(diskbb+gauss+smedge*powerlaw)const}, which resulted 
in $\chi^{2}_{red}$ of 64.55/77. 
The width of the Gaussian was fixed at 0.7 keV (varies between 0.62 keV and 0.88 keV)
which was obtained as an average to all the fits (except for the hard states). Systematic error of 
0.5\% was included in all the fits, for considering the uncertainities in the data. 
We obtained the parameters of disk temperature, disk normalization, photon index with the 
powerlaw norm and the line energy of the Gaussian line. The photon index of the HEXTE component was 
tied to that of the PCA component. Since the combined spectra showed a cut-off feature in the 
residuals at higher energies, we decided to include a \textit{highecut} model for considering 
this. The fits improved with $\chi^{2}_{red}$ = 57.57/75 for the model of \\
\textit{phabs*(diskbb+gauss+smedge*powerlaw*highecut)}. The F-test probability for 
inclusion of {\it highecut} model for 
this observation is 0.014, but for the observations during the rising phase of the outburst the 
F-test probability is of the order of 10$^{-28}$ to 10$^{-14}$ implying that inclusion of {\it highecut} 
model is statistically significant. We have considered the {\it highecut} model in order to understand 
the evolution of the cut-off and fold energy during the spectral states, although the F-test 
probability for inclusion of {\it highecut} during the soft-intermediate state is high. Figure 
\ref{ufspec-highecut} shows the unfolded spectrum for the fits using \textit{highecut} model for one 
of the observations. Hence, it was decided to consider \\
\textit{phabs*(diskbb+gauss+smedge*powerlaw*highecut)} as the final model, and similar method 
of analysis was followed for all the data sets during the entire outburst.

We also attempted to model the spectrum with \textit{compTT} \citep{Titarchuk1994} to take into 
account the parameters related to the non-thermal component of the spectra (Comptonized corona), 
and obtained $\chi^{2}_{red}$ = 72.59/78 for the same observation ID. Although the fit for the model 
\\ \emph{phabs*(diskbb+gauss+smedge*compTT)} was 
statistically good, the variations seen in the parameters were not able to explain the nature of 
the spectral evolution. We observed that as the source rise towards the peak, the optical depth 
varied from 1.5 to 0.01, and the seed 
photon temperature reduced from 2.5 to 0.49 which seems to be unrealistic, although the electron 
temperature increased from 29 keV to 239 keV. 

Recently \citealt{Fari2013} have performed the spectral analysis of the source in the context of a 
complex bulk motion comptonization model. In order to have a comparative study, we also applied the 
model of \\ \emph{wabs*(diskbb+gauss+bmc*highecut)const} to the same observation ID, and obtained 
$\chi^{2}$/dof = 93.83/77. A comparison with the analysis performed by \citealt{Fari2013} shows 
that the fit will probably improve if the BeppoSAX data is also included. In spite of the fact that 
we did not consider the BeppoSAX data, we studied this model for the different spectral states, and 
found that the variation of the parameters characterising the disk properties were similar to those 
shown in Figure 4 of \citealt{Fari2013}. We also found that the value of parameters related to 
the disk and the photon index obtained using the model we have applied and that used by 
\citealt{Fari2013} are within 1$\sigma$ error limit. 

Since MJD 51486.87 ($\sim$ $26^{th}$ day) to the end 
of the outburst, the HEXTE spectrum was found to be mostly background dominated 
(few cases spectra up to $\sim$ 50 keV). Hence, we decided to fit only the 3 - 25 keV PCA 
spectrum using {\it diskbb} and {\it powerlaw}. Better fits were obtained 
with the model of {\it phabs(diskbb+gauss+smedge*powerlaw)} with $\chi^{2}$/dof of 28.62/39 
(for observation ID: 40124-01-40-01).

It is to be noted that since we are considering RXTE data $\gtrsim$ 3 keV only, the value of inner 
disk temperature obtained from the fits will not be exact. \citealt{Mer2000} pointed out that the 
diskbb model parameters should not be considered as ideal, since the parameter of disk radius varies 
due to the change in spectral hardening factor (see also \citealt{Dunn2011b}).

The convolution model, {\it cflux} was included to find the flux contribution of the individual 
components, in the energy range of 3 - 150 keV. The error values for each parameter were obtained 
using the {\it err} command at a confidence interval of 90\%. We observe that the uncertainty 
in flux estimated using {\it cflux}, is not of large variation, except a few cases of the 
soft-intermediate state where a 5\% variation is observed. The variation between the values 
obtained by {\it cflux} and {\it flux} commands are found to be a maximum of 4\%.

\begin{figure}
\hspace{-0.5cm}
\includegraphics[width=8.8cm]{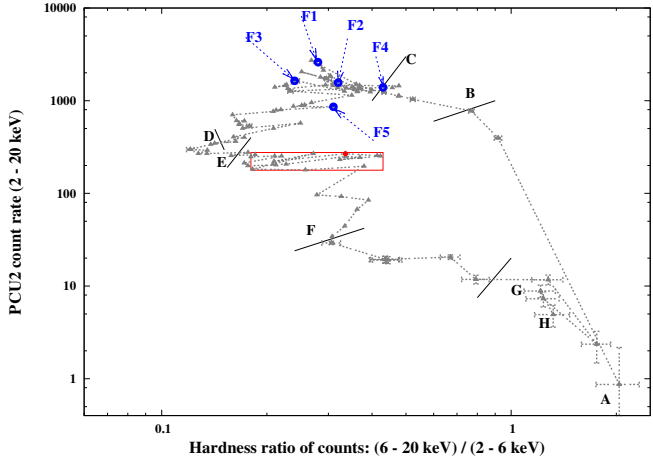}
\caption{Hardness Intensity Diagram (HID) for the source XTE J1859$+$226 during the 1999 outburst 
showing variation of background subtracted count rate with the color, which has been 
observed with RXTE/PCA (i.e., PCU2). 
The points marked A $\rightarrow$ H depict the start/state transitions/end of the outburst. 
The five flares (F1 $\rightarrow$ F5) observed along with the weak signature of secondary outburst 
phase (marked as rectangular box) are also shown. See text for details.}
\label{q-dia}
\end{figure}

\section {Results and Modeling}
In this section, we present the evolution of temporal and spectral properties 
(in broad spectral band of 3 - 150 keV) associated with different branches of HID of the source 
XTE J1859$+$226 to understand the accretion dynamics from a theoretical point of view. We estimate 
mass of the source by modeling the QPO evolution and also by correlation studies, and 
describe the X-ray features observed during the Radio flares and thereby investigate the disk-jet 
symbiosis. 

\subsection{Outburst evolution}
In this sub-section, we present the characteristics of the HID along with variation of X-ray and Radio 
flux. Figure \ref{rlc-pca} shows the intensity variations of the emission from the 
source, both in X-rays and radio. We observe that as the source rises from quiescence, 
the background subtracted X-ray Count rate (2 - 20 keV) rises from $\sim$ 1 
count/sec (MJD 51460.76 = $0^{th}$ day) to a peak of 2714 counts/sec (MJD 51467.58). 
Although for the first observation, the source count rate is very less with large error bars, this 
observation is important to follow the evolution of the source since the beginning of the outburst. 
Similar values were quoted by previous studies carried out by \citealt{HB2005} and 
\citealt{Casella2004}. The radio flux peaks (MJD 51467.9) at a value of 114 mJy 
resulting in peak flare F1, implying the release of matter in the form of jets \citep{Brocksopp2002}. 
Multiple ejections have been observed resulting in flare F2 on MJD 51475 at 19 mJy when X-ray 
intensity was 2045 counts/sec (on MJD 51474.28), 
flare F3 on MJD 51479.9 at 31 mJy during which the X-ray intensity was 1637 counts/sec, flare F4 at a 
peak flux of 5.2 mJy on MJD 51484.6 where the X-ray intensity is around 1391 counts/sec. The source 
count rate starts decreasing from 1444 counts/sec (after MJD 51483.94) to $\sim$ 4.9 counts/sec 
(MJD 51626.6 = $166^{th}$ day) and attains quiescence. 
While decaying, the source is observed to show a weak secondary emission (see \S 3.1.5) for around 
24 days with peak emission on $69^{th}$ day (MJD 51530.13) of the outburst. A fifth flare 
(F5) is reported by \citealt{Brocksopp2002} based on ASM lightcurve to occur on MJD 51487.6, but 
there were no radio observations on that day. 

Figure \ref{q-dia} shows the HID (see also \citealt{HB2005,Dunn2011a}) plotted using the 
background subtracted 
PCU2 count rate. The points where the five flares occurred are also marked in the figure (see also 
\citealt{FHB09}). 
The possible transitions which occurred, are labelled from A to H. In general, outbursting 
sources undergo state transitions from \textit{hard $\rightarrow$ hard-intermediate $\rightarrow$ 
soft-intermediate $\rightarrow$ soft $\rightarrow$ soft-intermediate $\rightarrow$hard-\\intermediate
 $\rightarrow$ hard} \citep{HB2005, TB2005, TB2010, Nandi2012} and finally into a 
quiescent state. 

\begin{figure}
\includegraphics[height=7cm,width=9.5cm]{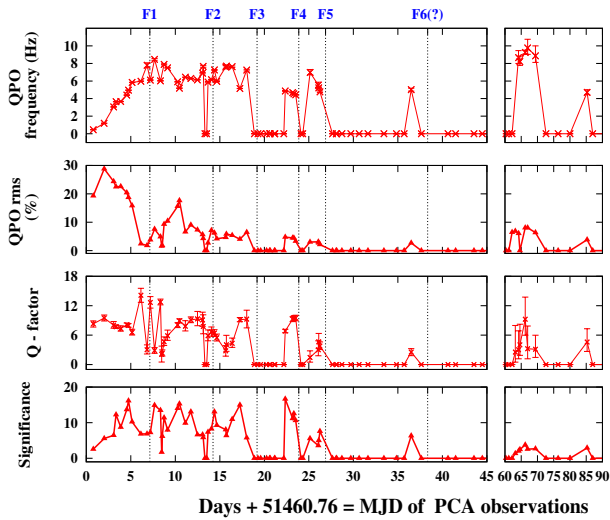}
\caption{Evolution of QPO frequency, QPO-rms, Q-factor and significance throughout the outburst of 
XTE J1859$+$226. Absence of QPOs during radio flares (ejections) are shown in the plot. Peak timings 
(in unit of days) of Radio flares are also marked as F1, F2, F3, F4 \& F5 
(see Table \ref{tabflares} for details).}
\label{qpo-evo}
\end{figure}

In Figure \ref{qpo-evo}, we show the variation of QPO frequency, QPO rms, the Q-factor and 
significance of the QPOs observed, over the entire outburst of the 
source XTE J1859$+$226. The QPO frequencies increase monotonically from 0.46 Hz to 6.09 Hz in the 
rising phase of the outburst, which is a quite natural phenomena as observed in other outbursting 
BH sources \citep{BH1990, TB2005,Nandi2012}. In the later phase of the 
outburst, the presence/absence of QPO frequencies 
are directly linked with Jet emission (see Figure \ref{qpo-evo} and \S 3.3) as well as with the 
`secondary' emission of the outburst (see Figure \ref{rlc-pca} and \S 3.1.5). 

Evolution of the parameters of the spectral fit, obtained from the analysis are shown in Figures 
\ref{spec-evo_F1234}, \ref{fold-ene} and \ref{spec-evo_F5-6}. Figure \ref{flux-evo_F123} depicts the 
variation of disk flux and powerlaw 
flux in the 3 - 20 keV range and also for the 20 - 150 keV powerlaw flux; and 
Figure \ref{flux-evo_ratio} shows the fraction of thermal flux contributing to the total 
3 - 20 keV flux, the contribution of thermal flux over non-thermal flux 
(3 - 20 keV) and also that of the flux ratio in 3 - 20 keV to 20 - 150 keV energy range (up to 
MJD 51486.82).

\subsubsection{Hard state (branch AB)}
During the first PCA observation (point A in Figure \ref{q-dia}), the PDS was dominated by broad-band 
noise. The next observation (on MJD 51461.54) has a red noise along with a QPO of 0.46 Hz 
(Q-factor = 8.31, significance = 2.6, rms amplitude = 19.3\%) which was not reported 
by \citealt{Casella2004}, and is marked as the first point in the top panel of Figure 
\ref{qpo-evo}, Figure \ref{qpo-pos} and Figure \ref{scal}. Figure \ref{qpo-evo} implies that the 
QPO frequency starts rising from 0.46 Hz to 1.19 Hz (type C QPOs; see \citealt{Casella2004} for 
details) with the power spectra being mostly dominated by red noise 
and the rms of PDS varies from 30\% to 24\%. This variation in total rms is similar to 
that observed in other BH sources like GX 339$-$4 (See \citealt{MD2010}). 
In this branch, the photon indices of the energy spectra rise from 
1.6 to 2.0, the disk temperature (T$_{in}$) varies in-between 1.01 keV to 0.83 keV 
(Figure \ref{spec-evo_F1234}) and hard X-ray flux dominates over the disk flux 
(Figure \ref{flux-evo_F123}); and the hard X-ray spectral component (i.e., fold energy) increased 
from 53 keV to 114 keV. 

\subsubsection{Hard-intermediate state (branch BC)}
In this branch, the power spectrum has a flat top noise along with 
QPOs and a powerlaw form at frequencies $>$ 10 Hz, with the total rms varying from 24\% 
to 16\% and the QPO frequency increasing from 3.05 Hz to 5.97 Hz (Type C; see 
\citealt{Casella2004}). The photon index increases from 2.0 to 2.4 and the disk temperature slowly 
increases from 0.75 keV to 0.85 keV. 
The disk flux is observed to be increasing as compared to the non-thermal powerlaw flux 
(Figure \ref{flux-evo_F123}), as well as the contribution of low energy to high energy 
powerlaw flux increases (Figure \ref{flux-evo_ratio}). 

Figure \ref{fold-ene} shows that during branch BC, the fold energy increased from 108.3 keV to 
201.2 keV. Thus we observe an increase in fold Energy from 53 keV to 201.2 keV, when the source 
moves through the branches AB and BC during the rising phase. \citealt{Fari2013} reported about an 
increase in the cut-off component during the transition from the hard to soft state. 

\subsubsection{Soft-intermediate state (branch CD)}

During the initial phase of the branch CD (from MJD 51468.4 to MJD 51473.82), the steepness of the 
red noise component starts decreasing and the power spectra has 
a flat top noise along with QPOs (both A and B types). QPOs of type C and also type B {\it Cathedral 
or twin QPOs} (\citealt{Casella2004}) are observed during this branch. QPOs observed after 
MJD 51484.276 have not been classified into any types (\citealt{Casella2004}). 
The total rms of the PDS is observed to increase from 7\% to 14\%, while during the end phase of the 
branch (after MJD 51484.87), the PDS has only broadband noise of amplitude 
6\% to 3\% rms, with or without QPOs along with a powerlaw form from 10 Hz to 64 Hz.

The fold energy is observed to vary randomly in between 64 keV and 180 keV, around an average of 
130 keV within 1$\sigma$ error bars (except few data points, which are within 1$\sigma$ to 2$\sigma$). 
The disk flux is seen to vary randomly, 
whereas the power-law flux decreases gradually as shown in Figure \ref{flux-evo_F123}. 
During the later phase of this state, the disk temperature slightly decreases from 0.85 keV to 
0.7 keV and photon index decreases from 2.4 to 2.0 (Figure \ref{spec-evo_F5-6}). Although 
\citealt{Fari2013} has reported a photon index of 1.8 during the 
period of MJD 51490 to 51501, we observed a photon index varying between 1.9 and 2.4 with disk flux 
contribution of $>$65\%. These parameters imply the characteristics of a soft-intermediate state 
during this period and not a hard state as suspected by \citealt{Fari2013}.

During this state, multiple ejection events have 
occurred and observed in Radio as flares \citep{Brocksopp2002}. 
\citealt{FHB09} has reported on the features of HID and temporal characteristics during a flare. We 
present detailed observational results of the spectral and temporal properties, during each of 
these flares in \S 3.3.

\begin{figure}
\hspace{-0.8cm}
\includegraphics[width=10cm]{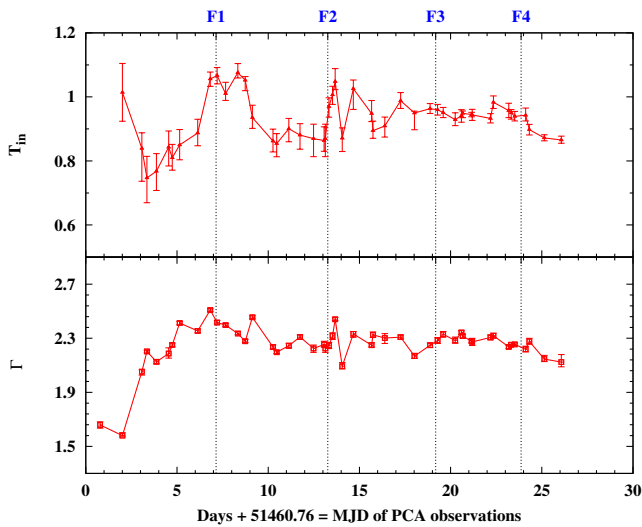}
\caption{Evolution of the spectral parameters, {\it disk temperature (T$_{in}$) and photon index 
($\Gamma$) of observations before flare 5 (F5) of the source 
XTE J1859$+$226.}}
\label{spec-evo_F1234}
\end{figure}

\begin{figure}
\hspace{-0.8cm}
\includegraphics[width=10cm]{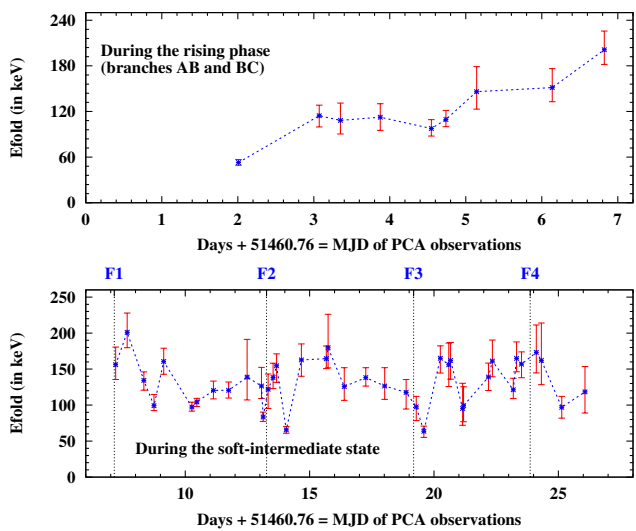}
\caption{Evolution of fold Energy (E$_{fold}$) during the rising phase (branches AB, BC; top panel) 
and soft-intermediate state (branch CD; bottom panel) of the HID for the source XTE J1859$+$226.}
\label{fold-ene}
\end{figure}

\begin{figure}
\hspace{-0.8cm}
\includegraphics[width=10cm]{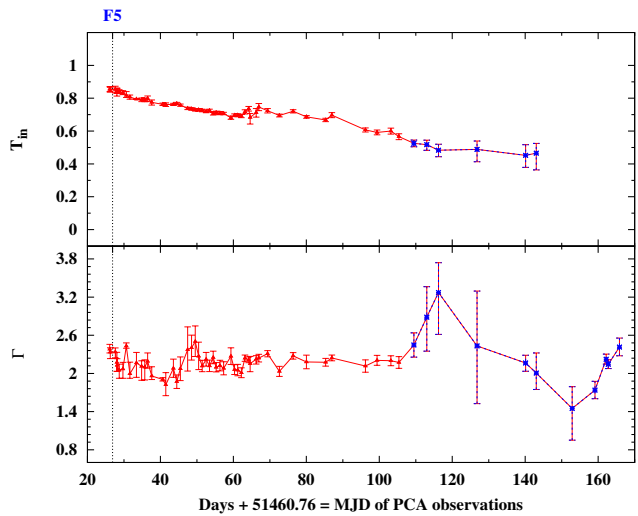}
\caption{Evolution of the spectral parameters (3 - 25 keV energy range), {\it disk temperature 
T$_{in}$, photon index $\Gamma$} for observations from flare 5 (F5) till the end of the 
outburst of XTE J1859$+$226. Note, spectral data is fitted up to 15 keV (around $120^{th}$ day) during 
the unusual variation of photon index.}
\label{spec-evo_F5-6}
\end{figure}

\subsubsection{Soft-state (branch DE)}

During the transition from branch CD, we find that a few of the observations (MJD 51504.31 to 
MJD 51510.29 in Figure \ref{rlc-pca}) show hardness ratio 
around 0.1 (Figure \ref{q-dia}) and total rms $\sim$ 1\% to 2\%, which suggest a short presence of soft 
state (see \citealt{MD2010}). Similar value of hardness ratio was quoted in Figure 19 of 
\citealt{Casella2004}, Figure 1 of \citealt{HB2005} and Figure A.1 of \citealt{Dunn2011a}, for this 
source. During these observations the thermal flux dominates the spectra (see top panel of 
Figure \ref{flux-evo_ratio}) and QPOs are also not observed. But 
the other spectral features during this phase seems not to imply the characteristics of the 
`canonical' soft state as suggested by \citealt{RM04}. We observe that as compared to branch CD 
(soft-intermediate state), the disk 
temperature reduces from 0.9 keV to 0.7 keV \textit{(to be noted that it is not the exact 
value, due to the limitations of RXTE and diskbb model \citep{Mer2000})} and the photon 
index remains $\sim$ 2.2 (spectra are harder compare to soft-intermediate state). These variations 
are very different 
from the observations of other BH sources where the disk temperature remains high above 1 keV, 
photon index is $>$2.5 during the soft state (See \S 4.3.5 and Figures 4.8 and 4.9 in \citealt{RM04};
\citealt{TB2006}).  
All these results deviate from the `canonical' description of soft spectral state observed 
in outbursting BH sources \citep{TB2005, TB2006, RM04, TB2010}.

\subsubsection{Soft-intermediate state (branch EF)}

In branch EF, no significant QPOs are observed (\citealt{Casella2004} has not classified the QPOs 
also), and the PDS has broadband noise of rms value $\sim$ 3\%. The source is in the declining phase 
and we observe that the disk and power-law flux start gradually decreasing 
(Figure \ref{flux-evo_F123}) without any signature of hard X-ray flux beyond 50 keV.

\begin{itemize}

\item The outburst profile (top panel of Figure \ref{rlc-pca}, see also Figure \ref{q-dia}) 
indicates that, during this phase there is an enhancement in X-ray flux (also observed by RXTE-ASM) 
and it gradually decreases to the quiescence. 
From the spectral analysis, we observed that there is an increase in the disk and powerlaw flux 
(Figure \ref{flux-evo_F123}) along with increase in disk temperature for $\sim$ 24 days i.e., 
from $63^{rd}$ to $87^{th}$ day of the outburst (Figure \ref{spec-evo_F5-6} and 
Figure \ref{rlc-pca}). Although the value of inner disk temperature (T$_{in}$) is less, we note 
that the variation in uncertainty of T$_{in}$ is less than the value of T$_{in}$ itself 
(see also \citealt{RM04}). The increase in disk and powerlaw flux can also be noted in Figure 14 
of \citealt{Dunn2011a}, after 
$\sim$ MJD 51530. Although such an increase in flux has been observed in other sources like GX 339$-$4 
\citep{Dunn2011a}, but that particular observation ($\sim$ MJD 52710) was associated with the 
hard-intermediate state \citep{Belloni2002,Motta2011}. 
The temporal analysis for the source 
XTE J1859$+$226 indicates weak signature of QPOs in the PDS, during this phase of branch EF 
for the days from $64^{th}$ to $85^{th}$ of the outburst as shown in Figure \ref{qpo-evo} (see also
\citealt{Casella2004}). Table \ref{secondary} which summarizes the details of temporal features 
during this phase, implies the presence of QPOs. QPOs of significance $\gtrsim$ 2.7 (with higher 
Q-factor) are observed during MJD 51527.004, MJD 51530.133 and MJD 51546.036. Other QPOs seem to
be weak (significance $\sim$ 2.5) in nature (except the QPO observed on MJD 51523.941 with 
significance of $\sim$ 2), but having Q-factor $\sim$ 3. Since the source
is in the declining phase of the outburst, these QPOs observed can be classified as type C*. Both the 
spectral and temporal properties seem to suggest that probably this phase/feature could be associated 
with a `secondary' emission, within the same outburst.   
\end{itemize}
 
As the source approaches the end phase of branch EF, there is a sudden increase in the photon 
index from 2.2 to 3.2 (spectral data fitted up to 15 keV), but with large errors (star points in 
Figure \ref{spec-evo_F5-6}), which turns out to be a puzzling nature.

\begin{table}
\addtolength{\tabcolsep}{-4.00pt}
\scriptsize
\centering
\caption{\label{secondary} Details of temporal characteristics during the `secondary' outburst}
\vskip 0.0cm
\begin{tabular}{ccccccc}
\hline \\
MJD & & Day & QPO frequency & Significance & QPO rms & Q-factor\\
    &  &   & (Hz)          &    of QPO      &(\%)    &\\
\hline\\
51523.941  &    & 63.18 &0      &0       &0     &0\\
51524.940  &	& 64.18	&8.66$^{+0.8}_{-0.6}$	&2.04    &6.87	&3.02$^{+3.8}_{-1.8}$\\
51525.353  &	& 64.59 &8.21$^{+0.6}_{-0.4}$ 	&2.48    &6.03	&4.02$^{+4.2}_{-2.1}$\\
51527.004  &	& 66.24 &9.25$^{+0.2}_{-0.1}$ 	&3.75    &8.02	&9.24$^{+4.5}_{-3.2}$\\
51527.736  &	& 66.97 &9.75$^{+0.9}_{-0.7}$ 	&2.56    &8.03	&3.23$^{+4.6}_{-2.0}$\\
51530.133  &	& 69.37 &8.86$^{+0.7}_{-1.1}$ 	&2.66    &6.32	&3.11$^{+2.8}_{-1.6}$\\
51546.036  &	& 85.27	&4.71$^{+0.2}_{-0.3}$	&2.90	 &3.82	&4.58$^{+2.7}_{-1.9}$\\
51547.766  &	& 87.00 &0	&0	 &0	&0\\	
\hline
\end{tabular}
\end{table}

It is observed that during both the soft-intermediate branches CD and EF, there are QPOs of types A, B, 
and a few C and C* (see also \citealt{Casella2004}). 

\subsubsection{Hard-intermediate state (branch FG)}

In branch FG, we do not observe any QPOs and the power spectra remain to have broadband noise 
with total amplitude (rms) varying from 7\% to 11\%.
The source is in the declining phase and the spectra is dominated by powerlaw component, although we 
find weak signature of thermal flux with reduced inner disk temperature.

\subsubsection{Hard state (branch GH)}

At the end of the outburst, the source finally moved towards the hard state, which has 
characteristics similar to branch AB. The amplitude (rms) of the power 
spectrum has increased to $\sim$25\% without any signature of QPOs. The energy spectra are mostly 
powerlaw in nature with photon index around 1.4 except during the last three observations which show 
a photon index of $\geq$ 2.1 although the source count rates 
are very less with large errors, and it was difficult to fit the spectra.

\begin{figure}
\includegraphics[width=9.5cm]{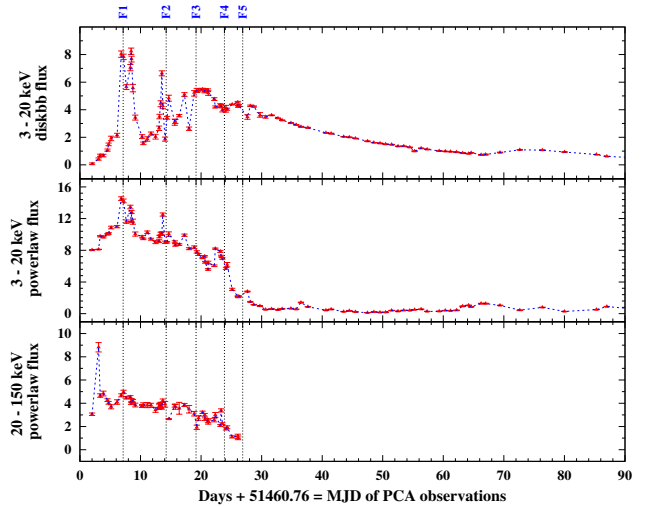}
\caption{Variation of the flux contribution by the disk component and powerlaw component in the 
energy range of 3 - 20 keV, and 20 - 150 keV powerlaw flux. Flux values are quoted in units of 
10$^{-9}$ erg cm$^{-2}$ s$^{-1}$.}
\label{flux-evo_F123}
\end{figure}

\begin{figure}
\includegraphics[width=9cm]{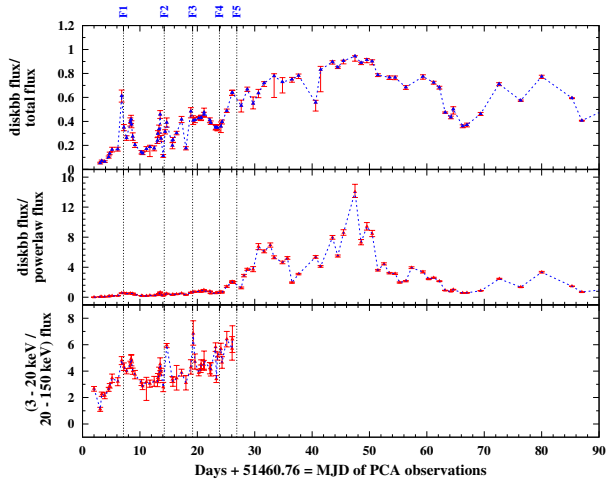}
\caption{Variation of the fraction of disk flux over the total flux, fraction of disk flux over 
powerlaw flux in the energy range of 3 - 20 keV, and of the ratio in 3 - 20 keV to 20 - 150 keV flux. 
Flux values are quoted in units of 10$^{-9}$ erg cm$^{-2}$ s$^{-1}$.}
\label{flux-evo_ratio}
\end{figure}

\subsection{Modeling of QPO frequency evolution and $\Gamma$ - QPO correlation}
\begin{itemize}
\item \textbf{QPO evolution} \\
The evolution of QPO frequencies during rising phase of XTE J1859$+$226 can be understood based on 
the Propagating Oscillatory Shock (POS) solution \citep{skc08,skc09, Nandi2012, Deb2013} of the TCAF. 
According to this model, the oscillating frequencies observed 
as QPOs are due to the movement of the shock surface (i.e., Comptonizing region). During the rising 
phase, the shock moves towards the BH and during the declining phase it moves away from the BH. The 
POS solution states that the QPO frequency is inversely proportional to the infall time scale, and 
the QPO frequency ($\nu_{QPO}$) can be obtained with a knowledge of the instantaneous shock location 
or vice-versa, and from the value of the compression ratio 
(R = $\rho_{+}$/$\rho_{-}$, where $\rho_{+}$ and $\rho_{-}$ are the 
densities in post and pre shock flows). 
According to \citealt{Ryu97}, \citealt{CM2000}, the frequency of the shock location 
is similar to the observed QPO frequency. Thus, the QPO frequency for an axisymmetric toroidal bounded 
system (i.e., shock surface; see Figure \ref{mtcaf}) can be written as

\begin{figure}
\includegraphics[width=8.8cm]{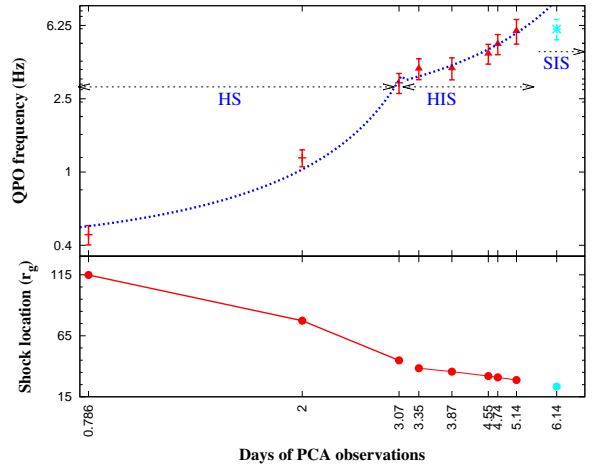}
\caption{Evolution of QPO frequencies with days, during the rising phase of the outburst of 
XTE J1859$+$226. The FWHM of the fitted Lorentzian of the QPO has been considered as the error bars. 
The curves show the fitted solution based on the POS model. The spectral states (HS: Hard state, 
HIS: hard-intermediate state, SIS: soft-intermediate state) has also been marked. 
The bottom panel shows the variation of the shock 
location (i.e., Comptonizing region; See also \S 4 for discussions on the same).}
\label{qpo-pos}
\end{figure}

\begin{eqnarray}
\nu_{QPO} = \frac{\nu_{s0}}{t_{infall}} = \frac{\nu_{s0}}{[2\pi Rr_{s}(r_{s}-1)^{1/2}]},
\end{eqnarray}

where $\nu_{s0}$ = c/r$_{g}$ is 
the inverse of the light crossing time of the BH, and r$_{g}$=2GM/c$^{2}$ for a BH of mass M. Since the 
shock will be drifting with time \citep{skc08,skc09}, the time dependent shock location can be 
expressed as 
\begin{eqnarray}
 r_{s}(t) = r_{s0} \pm v_{0}t/r_{g} , 
\end{eqnarray}

where r$_{s0}$ is the shock location at time t=0 and v$_{0}$ is the corresponding shock 
velocity \citep{CM2000,MSC96}. The `+' sign is used for an outgoing shock in 
the declining phase and the `-' sign is used for an inward shock during the rising phase. 

In Figure \ref{qpo-pos}, we show the evolution of QPO frequency with time (days) during the rising 
phase of XTE J1859$+$226, which we modeled using above equations of the POS solution. Fitted result 
allows to calculate the shock location and hence to estimate the size of the Comptonizing region. 
Similar attempts have been performed for GRO J1655$-$40, XTE J1550$-$564 
\citep{skc08,skc09}, GX339$-$4 \citep{Nandi2012} and H 1743$-$322 \citep{Deb2013}.

The first day of PCA observation is considered as t=0 sec. Figure \ref{qpo-pos} shows 
that the initial three observations of QPOs follow one trend while the rest shows another. Hence the 
evolution follows two different solutions for different parameters of shock location, velocity etc.. 
The FWHMs obtained from the fits to the Lorentzian feature of the QPOs have been considered as error 
bars. The fit to the initial three QPO frequencies gives the value of compression ratio as R=3.8, 
initial shock location r$_{s0}$ of 138.8 r$_{g}$ (t=0 sec) and the shock velocity v$_{s0}$ = 
8.5 ms$^{-1}$. The fit to the rest of 
the QPO frequencies during the rising phase gives R=3.2, r$_{s0}$=56.5 r$_{g}$, v$_{s0}$=1.5 ms$^{-1}$. 
The overall fit yields a mass of BH of 7.12 $\pm$ 0.54 M$_{\odot}$, with reduced $\chi^2$ of 0.69. 

In the final fitting, we excluded the last observation (as shown in cyan color) on 
MJD 51466.89, which is just before the first radio flare (F1). The variation of shock locations 
(i.e, the size of Comptonizing region) is shown in the bottom panel of the Figure. The oscillation of 
shock stalled at 21.2 r$_{g}$ and this gives the estimate of minimum size of the comptonizing region 
(before the flare F1).
From Figure \ref{qpo-pos}, it can be observed that there is a kink (see also \citealt{skc09}; 1998
outburst of XTE J1550$-$564) at the point where the two fitted functions meet. 

\item \textbf{$\Gamma$ - QPO correlation} \\
Based on the model developed by \citealt{TF04}, \citealt{Shap2007} found that a correlation 
exists between the QPO frequency and photon index for BH sources. They found that the correlation 
can be explained by the function 
\begin{eqnarray}
f(\nu) = A - DB ln (exp(\frac{\nu_{tr} - \nu}{D}) + 1 )
\end{eqnarray}
by scaling w.r.t a source which has similar evolution. 
The above correlation is linear with a slope for lower frequencies, and nearer to a 
particular frequency $\nu_{tr}$ it becomes constant. 
The parameter A denotes the value of the index saturation level, B is slope of the low-frequency part 
of the data and D controls how fast the transition occurs. 

In Figure \ref{scal}, we have shown the correlation of photon index and QPO frequency for the source 
XTE J1859$+$226. 

\begin{figure}
\includegraphics[width=9.0cm]{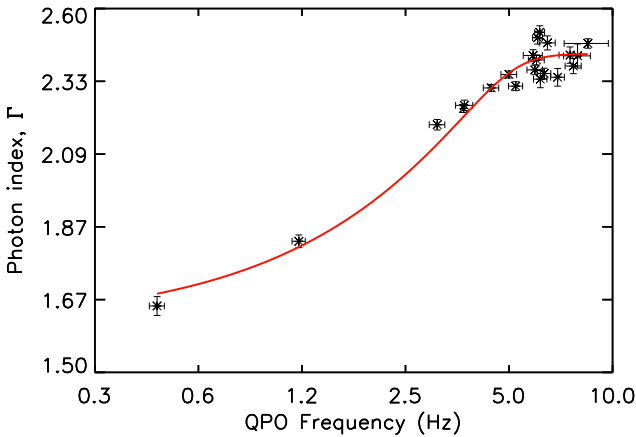}
\caption{Correlation of $\Gamma$ and QPO frequency during the rising phase of the outburst (before F2) 
of XTE J1859$+$226. The FWHM of the fitted Lorentzian of the QPO has been considered as the error 
bars.}
\label{scal}
\end{figure}

Similar studies was performed by \citealt{Shap2009} and the mass of the source XTE J1859$+$226 was 
estimated as 7.7 $\pm$ 1.3 M$_{\odot}$. In our analysis, we have considered the observation results of 
MJD 51461.54, which was not reported in \citealt{Shap2009}, and those during the rising phase of the 
outburst. We performed the scaling w.r.t GRO J1655$-$40, since the evolution of photon index of 
XTE J1859$+$226 saturates $\sim$ 2.4, which is closer to that of GRO J1655$-$40 at 
2.3, as mentioned in \citealt{Shap2007}. GRO J1655$-$40 is also a well studied source with a dynamical 
mass estimate of 6.3 $\pm$ 0.5 M$_{\odot}$. The value of D was fixed at 0.6 Hz. The fit was perfomed 
by applying Craig Markwardt's IDL routines of \emph{mpfit} \citep{Markmp}, considering the errors 
in $\Gamma$ and QPO frequencies.
The resultant fit parameters obtained were A1859 = 2.427 $\pm$ 0.008, B1859 = 0.168 
$\pm$ 0.008 Hz$^{-1}$ and $\nu_{tr}$ = 4.84 $\pm$ 0.18 Hz. Thus from the relation 
\begin{center}
M1859 = B1859 $\times$ (M1655/B1655), 
\end{center}

we obtained the mass of the source XTE J1859$+$226 as 7.96 $\pm$ 0.88 M$_{\odot}$, where 
B1655 = 0.133 $\pm$ 0.008 Hz$^{-1}$.  
\end{itemize}

\subsection{`Spectro-temporal' characteristics of the source during the flares}

In this section, we present the results based on the variations of temporal and spectral features
during radio flares i.e., when the jet ejections have taken place to understand the disk-jet
symbiosis. 
In Table \ref{tabflares}, we summarize the properties of the source during the flares. As 
mentioned in \S 2, we have performed an energy dependent study of the PDS (in the bands 2 - 6 keV, 
6 - 13 keV and 13 - 25 keV) of all the observations, in order to understand the nature of the X-ray
features during the flares.

\begin{table*}
\addtolength{\tabcolsep}{-4.00pt}
\scriptsize
\centering
\caption{\label{tabflares} Details of X-ray properties (in 2 - 25 keV band) of the observations 
during all the flares}
\vskip 0.0cm
\begin{tabular}{ccccccccccccc}
\hline \\
X-ray/Radio & X-ray Flux & MJD & Date & Time & Radio flux & \multicolumn{4}{c}{QPO} & $\Gamma$ & & T$_{in}$ \\\\
\cline{7-10}\\
(ID/Flare)& (cts/sec) & & &  & (mJy) & frequency(Hz) & rms(\%)  & sig. & type & & & (keV)\\
   
\hline\\
40124-01-11-00$^{\bf b}$ & 1505 & 51466.896& 1999-10-15&21:30:08 & & 5.97$\pm{0.02}$ & 15.4 & 6.83 & C & 2.35$^{+0.01}_{-0.02}$ & & 0.88$^{+0.04}_{-0.05}$ \\\\
40124-01-12-00 & 2714 & 51467.581& 1999-10-16&13:57:20 & & 0 & 0 & 0 & - & 2.51$\pm{0.01}$  & & 1.05$\pm{0.02}$\\
 &  & & & (7 hrs)$^{\bf a}$ & \\
\bf F1& & 51467.904 & 1999-10-16 & 21:41:45.6 & 114 & \\\\
40124-01-13-00$^{\bf c}$ & 2597 & 51467.961 & 1999-10-16&23:04:16 & & 6.1$\pm{0.02}$ & 3.8 & 7.24 & B & 2.42$^{+0.02}_{-0.01}$  & & 1.07$\pm{0.02}$ \\\\
\hline\\
40124-01-23-01$^{\bf b}$ & 1558 & 51473.890& 1999-10-22&21:22:08 & & 7.6$^{+0.11}_{-0.09}$ & 4.3 & 5.82 & C* &2.22$^{+0.02}_{-0.03}$ & & 0.87$\pm{0.05}$\\\\
40124-01-15-02 & 1726 & 51474.087& 1999-10-23&02:06:40 & & 7.3$\pm{0.3}$ $^{\dagger}$ & 5.4 & 5.75 & C* & 2.25$\pm{0.02}$ & & 0.97$\pm{0.03}$\\\\
40124-01-15-03 & 2045 & 51474.287& 1999-10-23&06:54:40 & & 0 & 0 & 0 & - & 2.32$\pm{0.03}$ & & 1.00$\pm{0.03}$ \\\\
40124-01-24-00$^{\bf c}$ & 1984 & 51474.429& 1999-10-23&10:18:08 & & 5.87$\pm{0.07}$ & 2.7 & 7.36 & B & 2.44$^{+0.02}_{-0.01}$ & & 1.04$^{+0.04}_{-0.02}$ \\\\
40124-01-25-00 & 1248 & 51474.820& 1999-10-23&19:41:20 & & 6.14$\pm{0.04}$ & 7.1 & 8.27 & C & 2.09$^{+0.03}_{-0.02}$ & & 0.87$^{+0.03}_{-0.05}$ \\
&  & & & (6 hrs)$^{\bf a}$ & \\
\bf F2& & 51475.0 & 1999-10-24 & 00:00:00 & 19 &  \\\\
\hline\\
40124-01-31-00$^{\bf b}$ & 1262 & 51478.777& 1999-10-27&18:39:28 & & 7.24$^{+0.08}_{-0.09}$ & 6.4 & 5.76 & C* & 2.16$^{+0.02}_{-0.01}$ & & 0.94$^{+0.04}_{-0.05}$\\\\\\
40124-01-32-00 & 1637 & 51479.635& 1999-10-28&15:15:28 & & 0 & 0 & 0 & - & 2.24$^{+0.02}_{-0.01}$ & & 0.96$^{+0.01}_{-0.02}$ \\
 &  & & & (7 hrs)$^{\bf a}$ & \\
\bf F3 & & 51479.940 & 1999-10-28 & 22:45:07.2 & 31&  \\\\
40124-01-33-01 & 1625 & 51480.044& 1999-10-29&01:04:32 & & 0 & 0 & 0 & - & 2.28$\pm{0.02}$ & & 0.96$\pm{0.01}$ \\\\
40124-01-36-00$^{\bf c}$ & 1474 & 51483.106& 1999-11-01&02:33:20 & & 4.87$\pm{0.04}$ & 4.7 & 16.6 & B & 2.32$\pm{0.01}$ & & 0.98$\pm{0.02}$ \\\\
\hline\\
40124-01-37-02$^{\bf b}$ & 1391 & 51484.275& 1999-11-02&06:37:20 & & 4.43$\pm{0.02}$ & 3.3 & 10.6 & B & 2.25$\pm{0.01}$ & &  0.94$\pm{0.01}$\\
 &  & & & (8 hrs)$^{\bf a}$ & \\

\bf F4 & & 51484.625 & 1999-11-02 & 15:00:35.8 & 5.2 & \\\\
40124-01-38-00$^{\bf b}$ & 1266 & 51484.872& 1999-11-02&20:56:16 & & 0 & 0 & 0 & - & 2.21$^{+0.03}_{-0.02}$ & &  0.94$\pm{0.02}$\\\\
40124-01-39-00$^{\bf c}$ & 952 & 51485.874 & 1999-11-03&20:59:44 & & 6.97$^{+0.4}_{-0.2}$ & 3 & 5.56 & C* $^{\bf e}$ & 2.14$^{+0.03}_{-0.02}$ & & 0.87$\pm{0.01}$ \\\\
\hline\\
40124-01-41-00$^{\bf b}$ & 870 & 51487.009& 1999-11-05&00:13:36 & & 4.77$^{+0.07}_{-0.08}$ & 2.1 & 7.59 & C* $^{\bf e}$ & 2.34$\pm{0.1}$ & & 0.85$^{+0.01}_{-0.02}$ \\
 &  & & & (15 hrs)$^{\bf a}$ & \\
\bf F5$^{\bf d}$ & & 51487.63 & 1999-11-05 & 15:15:23.1 & - & \\\\
40124-01-42-00 & 857 & 51488.41 & 1999-11-06& 09:49:36 & & 7.39$^{+1.1}_{-0.7}$ $^{\dagger}$ & 3.1 & 1.3 & C* $^{\bf e}$ & 2.33$^{+0.07}_{-0.09}$ & & 0.84$\pm{0.02}$\\\\
	       &     & 51488.48 & 1999-11-06 & 11:31:28	& & 7.34$^{0.39}_{-0.34}$ & 3.1 & 4.5 & C* $^{\bf e}$ & & & \\\\ 
40124-01-43-00 & 765 & 51489.47 & 1999-11-07 & 11:27:28 & & 0 & 0 & 0 & - & 2.05$^{+0.09}_{-0.07}$ & & 0.85$\pm{0.02}$ \\\\ 
40124-01-49-00 & 502 & 51496.462& 1999-11-14& 11:06:24& & 0 & 0 & 0 & - & 2.10$\pm{0.02}$ & & 0.79$^{+0.01}_{-0.02}$ \\\\\\

40124-01-49-01$^{\bf c,b}$ & 571 & 51497.254& 1999-11-15& 06:06:40& & 5.02$\pm{0.08}$ & 2.7 & 6.32 & C* $^{\bf e}$ & 2.18$^{+0.14}_{-0.1}$ & & 0.79$\pm{0.02}$ \\\\

40124-01-50-01 & 502 & 51498.390& 1999-11-16& 09:22:24& & 0 & 0 & 0 & - & 1.96$^{+0.05}_{-0.06}$ & &0.77$^{+0.01}_{-0.02}$\\
 &  & & & (16 hrs)$^{\bf a}$ & \\
\bf F6 (?) & & 51499.100 & 1999-11-17 & 02:24:00.0 & 7.9 &  \\\\
40124-01-51-00 & 406 & 51501.384& 1999-11-19& 09:13:36& & 0 & 0 & 0 & - & 1.90$^{+0.03}_{-0.02}$ & &0.76$\pm{0.01}$\\

\hline
\end{tabular}
\\
Absence of QPO during F1 (in 2 - 5 keV and 13 - 25 keV band), whereas QPOs were absent in 2 - 25 keV 
during all other flares. {\bf a} - minimum duration of X-ray observation before flare;  
{\bf b} - observation of last QPO before flare; {\bf c} - observation when QPO reappeared; 
{\bf d} - No Radio observations, flare day based on ASM light curve \citep{Brocksopp2002}, radio 
flare might have occurred after this (see Figure \ref{rlc-pca}); {\bf e} - Not classified 
in \citealt{Casella2004}; $\dagger$ - weak indication
\end{table*}

\subsubsection{FLARE - I (F1)}
 
\begin{figure}
\includegraphics[width=9cm]{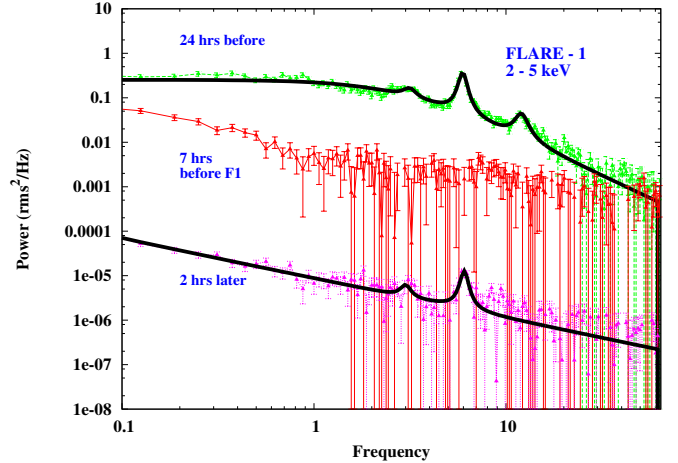}
\caption{Variation of the PDS implying that the QPO is not observed in the energy range of 2 - 5 keV, 
during the first flare (F1). The power shown in y-axis has been scaled by a factor of 100, 1 and 0.1, 
from the top to bottom. Similar variations also observed in the power spectra of 13 - 25 keV 
band.}
\label{F1-2-5-keV}
\end{figure}

The X-ray observation $\sim$ 24 hrs before (MJD 51466.896) the radio flare peak on MJD 51467.9, 
shows the presence of a type C QPO of 5.97 Hz. For the next observation (MJD 51467.581) which 
is 7 hrs before the flare peak we performed energy dependent study in narrower bands (2 - 4 keV, 2 - 5 
keV, 2 - 6 keV etc.) and found that the power spectra for 2 - 5 keV and 13 - 25 keV bands does not 
show any signature of QPO. A broad 7.79 Hz type A QPO is observed, only above 5 keV as evident in the 
6 - 13 keV PDS with rms amplitude of 3\%, which is typical for type A QPOs. So, in the overall 
range of 2 - 25 keV, a broad QPO of 7.79 Hz is observed (see also \citealt{Casella2004}).  
The next observation which is 2 hrs 
after the flare peak, shows a type B QPO of 6.1 Hz in all the energy bands. We 
also observe that the total rms of the PDS reduces (see also \citealt{FHB09}) to a value of 2.7\% when 
the QPO is not observed. Since the QPO was not observed in the soft band of 
2 - 5 keV, we have shown the variation in the PDS over 2 - 5 keV in Figure \ref{F1-2-5-keV}, for 
observations before and after the flare peak. Similar variations in the power spectra of
13 - 25 keV band also observed (not shown in Figure).
 
It is seen that, before the flare occurred i.e., when the QPO was observed, the hard photons 
slightly lag behind the soft photons. 7 hrs before the flare peak, the 2 - 5 keV soft photons are 
lagging compared to 5 - 25 keV hard photons. When the QPO is observed again in the PDS, the hard 
photons are observed to lag the soft photons. 

We also note that before the flare peak, the photon index increased to 2.51, 
disk temperature increased from 0.88 keV to 1.06 keV (Table \ref{tabflares} and 
Figure \ref{spec-evo_F1234}) and the fold energy of 
the electrons increased from 151 keV to 201 keV. The disk flux increased from 
2.14 $\times$ 10$^{-9}$ erg cm$^{-2}$ s$^{-1}$ to 8.03 $\times$ 10$^{-9}$ erg cm$^{-2}$ s$^{-1}$ 
(Figure \ref{flux-evo_F123}) and dominated over the powerlaw flux (3 - 20 keV) 
by a factor of 0.55. The 3 - 20 keV flux is found to be more than the 20 - 150 keV flux, 
by a factor of 5 (Figure \ref{flux-evo_ratio}), suggesting the soft nature of the spectrum. This 
implies that the emission is disk dominated. 

After the flare, the photon index decreased to 2.42 and disk flux
reduced to 7.83 $\times$ 10$^{-9}$ erg cm$^{-2}$ s$^{-1}$. This implies the hard nature 
of the spectrum although the disk temperature was around 1.07 keV and the fold energy had 
reduced to 156 keV.

Thus the broad QPO seen (just 7 hrs before the flare peak) in the 6 - 13 keV band while not 
in 2 - 5 keV and 13 - 25 keV, implies that the QPO is `partially' observed (see also 
\citealt{Mark2001}). We could not observe a `complete absence of the QPO' (i.e., in 2 - 25 keV band)
due to lack of continuous X-ray observations during the flare and just before/after. 
The spectrum also softens as compared to the spectrum of the observation before the flare.

\begin{figure}
\includegraphics[width=9cm]{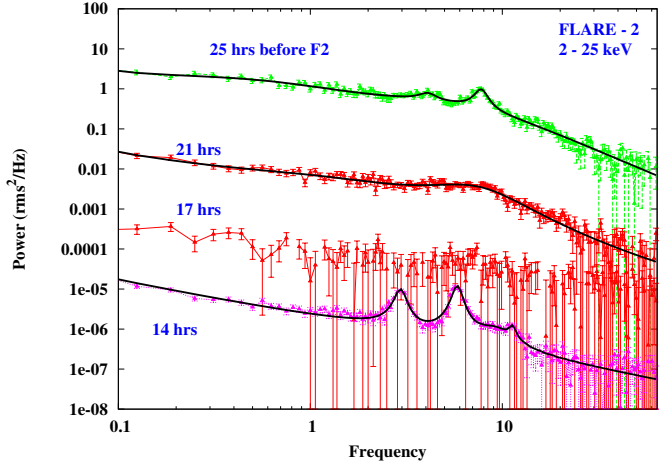}
\caption{Evolution of the PDS in the energy range of 2 - 25 keV during the 2nd flare (F2). 
Absence of QPO is seen $\sim$ 17 hrs before the radio flare peak of F2. The power
spectra are scaled by factor of 1000, 10, 1, 0.01 from top to bottom.}
\label{pds-F2}
\end{figure}

\subsubsection{FLARE - II (F2)}

Radio observation shows that the flare F2, peaked on MJD 51475.0 with a flux of 19 mJy. 
25 hrs earlier (MJD 51473.89) to the peak of F2, we observe type C* QPOs in the PDS. 
The observations $\sim$21 hrs before (MJD 51474.087) peak of F2 shows a weak signature 
of QPO with an increase in frequency (7.3 Hz with Q-factor of 1.2, significance of 5.75). 4 hrs 
later (MJD 51474.287) the PDS of F2 
shows complete absence of QPOs and the total rms reduces to 3.02\% (see also \citealt{FHB09}). 
\textit{Cathedral B-type} QPOs are observed at frequencies 3 Hz \& 6.1 Hz in the next 
X-ray observation which is 14 hrs before (MJD 51474.429) peak of F2. The evolution of power 
spectra is shown in Figure \ref{pds-F2}, which clearly indicates the `absence' of QPO during the 
flare, but QPO is observed again before the flare peak (see \S 4 for discussion).
 
We find that when the QPO is not significant in the power spectra, the soft photons are 
lagging. 17 hrs before peak of flare F2 where there is a complete absence of QPO, no lag of soft 
to hard photons 
is seen. When the QPO is observed again, $\sim$14 hrs before peak of F2, the hard 
photons are observed to lag behind the soft photons. 

The corresponding softening of the spectra is also observed in the variation of the 
spectral parameters. When the QPO was not present (over MJD 51473.89 to MJD 51474.287), we noted that 
the disk temperature increases from 0.87 keV to 1 keV, and the photon index also increased from 
2.22 to 2.32. The disk flux increased from
2 $\times$ 10$^{-9}$ erg cm$^{-2}$ s$^{-1}$ to 7 $\times$ 10$^{-9}$ erg cm$^{-2}$ s$^{-1}$ 
(Figure \ref{flux-evo_F123}), but the fold energy increased from 83 keV to 138 keV. 
The ratio of disk flux over powerlaw flux increased from 0.43 to 0.65 with a significant
decrease in hard X-ray flux (Figure \ref{flux-evo_ratio}). When the QPO was observed again, 
the photon index further increased to 2.44, with the disk temperature $\sim$ 1 keV only. The disk 
flux reduced to $\sim$ 4.28 $\times$ 10$^{-9}$ erg cm$^{-2}$ s$^{-1}$, whereas the fold energy 
increased to 154 keV. 
This suggest that, the nature of variation in spectral features along with the temporal properties 
(i.e., QPO features) during F2, is different from that of flare F1 and hence disk-jet symbiosis 
could be further complex (see \S 4 for discussions).

\subsubsection{FLARE - III (F3)}

\begin{figure}
\includegraphics[width=9cm]{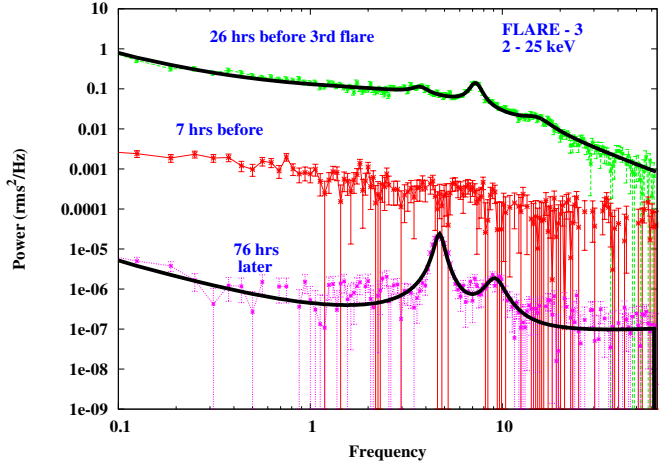}
\caption{Evolution of the PDS in the energy range of 2 - 25 keV during the flare F3. 
QPOs are absent $\sim$ 7 hrs before the radio flare peak of F3. The power
spectra are scaled by factor of 100, 10, 0.01 from top to bottom.}
\label{pds-F3}
\end{figure}

On MJD 51478.777, a 7.2 Hz type C* QPO is observed. Radio observations indicate that multiple 
flares (see Figure \ref{rlc-pca}) have occurred on MJD 51479.94 within 
$\sim$ 40 hrs of gap with flux values of 31 mJy \& 24 mJy during the period when F3 occurred. The 
X-ray observation which is around 7 hrs before F3, does not show any signature of QPO and the PDS 
is dominated by broad-band noise component (Figure \ref{pds-F3}). The subsequent X-ray 
observations does not show QPOs and the total rms decreased (see also \citealt{FHB09}) to 2.4\%. 
QPOs are observed almost after 3 days at a frequency of 4.8 Hz (type-B). 

\begin{figure}
\includegraphics[width=9cm]{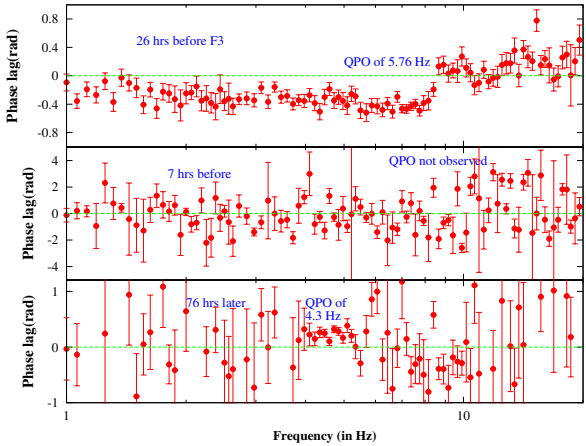}
\caption{The variation in phase lag between 2 - 6 keV and 6 - 25 keV, during the occurrence of 
flare F3. No phase lag is observed when QPO is not observed in the PDS.}
\label{lag-F3}
\end{figure}

We find an indication of soft photon lagging just before the flare peak, whereas when  
the QPO is not at all observed ($\sim$ 3 days), no lag is observed. The hard photons are 
observed to be lagging, when the QPO is observed again after 3 days (see Figure \ref{lag-F3}). 

During F3, disk temperature and photon index increases slightly, whereas the soft flux 
increased from 2.2 $\times$ 10$^{-9}$ erg cm$^{-2}$ s$^{-1}$ to 5.2 $\times$ 10$^{-9}$ 
erg cm$^{-2}$ s$^{-1}$ (Figure \ref{flux-evo_F123}), which implies the softening of the 
spectra. The fold energy decreased from 126 keV to 97 keV. We also note that during the 
observations 
where there were no QPOs, disk flux dominates over the powerlaw flux by a factor of 0.6 to 0.9 with
a significant decrease in hard X-ray flux (see bottom panel of Figure \ref{flux-evo_ratio}). 
At the same time, the powerlaw index and fold energy varies, but the disk temperature remains almost 
constant (Figure \ref{spec-evo_F1234}). 

The QPOs observed before and after the ejections during the flares F1, F2 and F3 are of 
type C/C* and type B respectively (see Table \ref{tabflares}).

\subsubsection{FLARE - IV (F4)}

Around 8 hrs before (i.e. on MJD 51484.27) the peak of next flare F4, a type B QPO is 
observed. The subsequent observations (on MJD 51484.87 and MJD 51485.07) does not show 
any signature of QPOs (Figure \ref{qpo-evo}), except broadband noise in the PDS (power spectral 
evolution not shown here) with decreased total rms of 2.5\% which is consistent with \citealt{FHB09}. 
It is also clear from Figure \ref{rlc-pca} that a flare has occurred with a 
flux of 5.2 mJy (MJD 51484.625). 
\citealt{Brocksopp2002} reported the time of flare based on X-ray observation from ASM. 
6 hrs after the peak of F4 (MJD 51484.87), QPOs are completely absent in the power spectra 
and are observed only after 30 hrs (on MJD 51485.874) at a frequency of 6.97 Hz 
(type C* with less rms and 
less Q-factor; see \citealt{Casella2004} for details of type C* QPOs).

During the flare F4, soft photons are observed to be lagging, while in flares F2 and F3, no lag of
soft to hard photons was observed.

It is also seen from Figure \ref{spec-evo_F1234} and Table \ref{tabflares} that during the flare 
peak, the spectral index varied around 2.2. The disk temperature remained around 0.9 keV, whereas 
the fold energy 
varied around 150 keV (Figure \ref{spec-evo_F1234}). The ratio 
of low energy to high energy flux increases with a factor of $\sim$ 6 (lower panel of 
Figure \ref{flux-evo_ratio}), implying the nature of spectral softening. The disk 
flux dominated over the non-thermal flux by a factor of 0.57 to 0.71, as seen in 
Figure \ref{flux-evo_ratio}. When the QPO was observed again, the photon index decreased 
to 2.14, with inner disk temperature of $\sim$ 0.87 keV, and decreased fold energy of 96 keV.

\subsubsection{FLARE - V (F5)}

\begin{figure}
\hspace{-0.5cm}
\includegraphics[width=9cm]{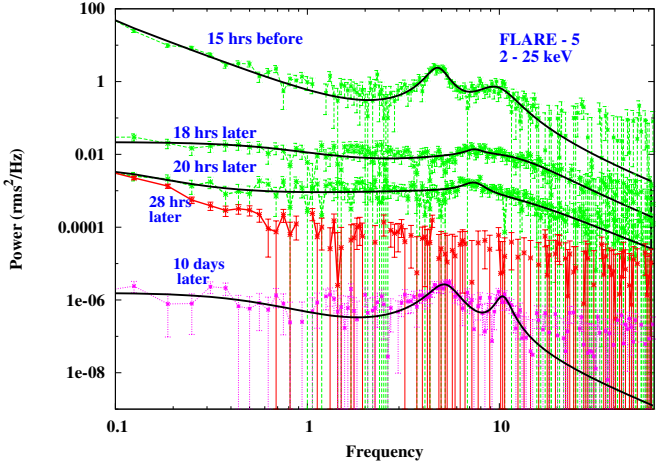}
\caption{Evolution of the PDS in the energy range of 2 - 25 keV during the flare F5. Scaling 
factors of 10000, 100, 10, 1 and 0.01 are applied for the power spectra from top to bottom.}
\label{pds-F5}
\end{figure}

We observe a 4.7 Hz QPO in the PDS of MJD 51487.01. Although the Q-factor is less, these QPOs 
can be considered as of Type C* class based on the QPO frequency and amplitude. 
The next X-ray observation (MJD 51488.41) indicates the presence of a weak QPO at 7.39 Hz (type C*) in 
the PDS of MJD 51488.41 and another type C* QPO at 7.34 Hz on MJD 51488.48. During the observation 
on MJD 51489.47 QPOs are not observed in the power spectra (see Figure \ref{pds-F5}). 
\citealt{Brocksopp2002} has reported a ejection/jet (radio flare), to have occurred on MJD 51487.63, 
based on the X-ray observations by ASM (see also \citealt{FHB09}). But there were no radio 
observations (see bottom panel of Figure \ref{rlc-pca}) during this time. The weak presence of QPO 
on MJD 51488.41 and MJD 51488.48, was followed by a complete absence of QPO in the power spectra of 
MJD 51489.47 along with softening of the spectra. This indicates that the fifth radio flare would 
have probably occurred after the time reported by \citealt{Brocksopp2002}. This is also implied by 
the detection of a 2.4 mJy radio flux on MJD 51490.4 (see Figure \ref{rlc-pca}). QPOs (type C*) are 
observed only after $\sim$10 days (MJD 51497.25).

Thus during F5 we observe that the QPO observed before and after the flare is of type C*,
whereas during flare F4 the evolution of QPO is from type B to type C*.
These features are not similar to that observed during the flares F1, F2 and F3.

The variation in spectral parameters does not show any significant change during the flare F5. The
photon index is seen to vary in between 2.38 and 2.33, while the disk temperature remains almost
constant around 0.79. 
Although the disk flux decreased from 4.24 $\times$ 10$^{-9}$ erg cm$^{-2}$ s$^{-1}$ to $\sim$
3.48 $\times$ 10$^{-9}$ erg cm$^{-2}$ s$^{-1}$, the ratio of disk to power-law flux is observed to 
increase by a factor of 3 to 6.7 (top panel of Figure \ref{flux-evo_ratio}) along with 
significant decrease of hard X-rays photons (beyond 50 keV), implying the softening nature of 
the spectrum. When the QPO is observed again, the photon index has reduced to 
2.18 with the disk temperature of $\sim$0.79 keV. The disk flux value is observed to 
dominate over the non-thermal flux, but with a lesser factor of 1.9. 

We observe a signature of QPO (type C* with lesser Q-factor) at 5.02 Hz (on MJD 51497.25), 
after 10 days of flare F5. The next observation (on MJD 51498.39) does not show any QPOs 
(Figure \ref{qpo-evo}) in the PDS, except a broad 
band noise. The total rms of the PDS reduces from 4.02\% to 1.6\%. Radio observations also show that the 
flux decreases to $\sim$ 1 mJy and later the source becomes non-detectable. We find that, there is a 
radio brightening (see \citealt{Brocksopp2002}) on MJD 51499.10 (after 16 hrs of X-ray obs.) of flux of 
7.9 mJy (Figure \ref{rlc-pca}), implying the possible signature of 
occurrence of a flare (F6?) that could have triggered on or before this time.

Since the evolution of the different types of QPOs during the ejections seems to be complex, we attempt 
to understand if any correlation exists between the QPO frequencies of different QPO types, with the 
thermal and non-thermal flux. The same has been presented in the following sub-section. 

\subsection{Relation between QPO frequency and flux variation}

\begin{figure}
\includegraphics[width=9.5cm]{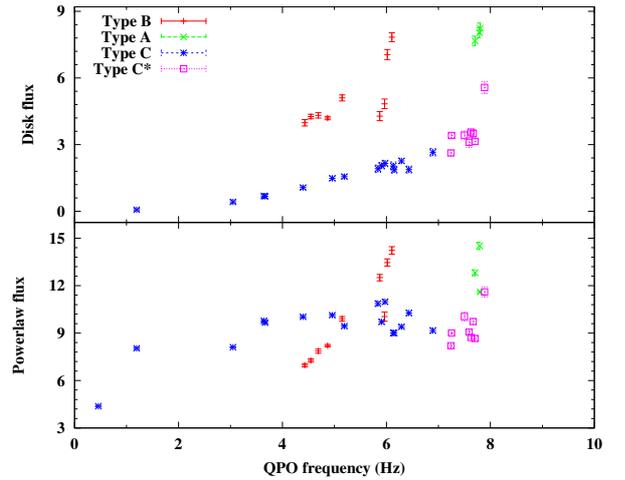}
\caption{Variation of the disk flux and powerlaw flux with increase in QPO frequency. The flux 
values are quoted in units of 10$^{-9}$ erg cm$^{-2}$ s$^{-1}$.}
\label{flux-qpo}
\end{figure}

In Figure \ref{flux-qpo}, we show the variation of disk and powerlaw flux as a function of QPO 
frequency for the observations in hard state (AB), hard-intermediate state (BC) and up to MJD 51484.27 
during flare F4 in the soft-intermediate (branch CD) of HID (see Figure \ref{rlc-pca} \& \ref{q-dia}), so as to consider the 
types of QPOs which have been already classified in \citealt{Casella2004}. We observed that during 
the rising phase (hard and hard-intermediate states) the QPO frequency of type-C increases as both the 
disk and powerlaw flux increases, implying a positive correlation. During 
the soft-intermediate state (branch CD) the type B QPOs are observed to be correlated with the 
powerlaw (non-thermal) flux but not with the disk flux. A few type C QPOs detected during the 
soft-intermediate state (branch CD) are observed not to show any correlation with the powerlaw flux. 
Figure \ref{flux-qpo} also shows that a few of the type C* QPOs have a correlation with 
the disk flux but are not at all correlated with the powerlaw flux as indicated by the random 
variation. Only very few type A QPOs are observed, and hence it is not possible to find any 
correlation with disk or powerlaw flux.

\section{Discussions and concluding remarks}

The evolution of temporal and spectral properties of the source XTE J1859$+$226 during the 1999
outburst suggest that the source does evolve via different spectral states throughout the outburst. 
Detailed analysis during the multiple ejections (i.e., radio flares) reveals that the 
ejections are possibly associated with the Comptonized corona, and hence
the sudden `appearance' of such radio flares or `jets' could be because of some instability happening 
in the Comptonized corona. Although several attempts have
been made to understand the evolution of `spectro-temporal' characteristics as well as the ejection 
events observed in outbursting BH sources, till date, there is no comprehensive picture for the 
1999 outburst of XTE J1859$+$226.

In this paper, we attempted to understand the possible accretion dynamics associated with the 
evolution of various X-ray features of XTE J1859$+$226, in the context of
Two Component Advective Flow (TCAF) model (\citealt{ST95, GC2013}). According to TCAF, the
Compton cloud (formed in shocked accretion phase) is nothing but the sub-Keplerian matter dominated 
CENtrifugal pressure supported BOundary Layer (i.e., CENBOL), which
intercepts soft photons, mostly originated from the Keplerian disk, and the CENBOL up-scatters them 
via inverse Comptonization to produce high energy photons. The CENBOL forms due to shock 
transition in the flow and its surface may oscillate due 
to the oscillation of shock, which could be responsible for the generation of QPOs  
(\citealt{MSC96, Ryu97, CM2000, skc04, Okuda2007, LRC2011}, see references therein).   

During the rising phase, we observed a monotonic increase of QPO frequency (C type) from 0.46 Hz to 
5.84 Hz in both hard and hard-intermediate states. This happens because, the shock location propagates 
towards the BH and the size of the shock 
formed region i.e., the CENBOL surface or region of Comptonization, (which is oscillating) reduces, 
and hence the QPO frequency increases (\citealt{skc08, skc09, Nandi2012}). We 
estimated the shock location (size of CENBOL) based on the POS solution (see \S 3.3). 
During this period, the instantaneous shock location (see Figure \ref{qpo-pos}) and hence the 
`Compton cloud/CENBOL' varies from 114.7 r$_g$ to 28.7 r$_g$. Using the same solution, we also 
estimated the size of the CENBOL region, before and after the flare. 

During the soft-intermediate state, the QPO frequency (all types of QPOs) varies in between 6 Hz 
to 8 Hz, and this implies that the size of the corona/CENBOL remains almost constant 
(i.e., shock location varies between 21.1 r$_g$ to 17.5 r$_g$). We also found that QPOs in the 
soft-intermediate state are sporadic in nature i.e. OFF (QPO not observed) and ON
(QPO observed again), which could be linked with jet ejection (see Figure \ref{qpo-evo} and below for
discussion). 
During the declining phase of soft-intermediate (except during the `secondary' emission of outburst), 
hard-intermediate and hard states, QPOs are hardly observable, whereas for other BH sources like 
GX 339$-$4, H 1743$-$322, XTE J1550$-$564 QPOs are observed during all the spectral states (except 
soft states) \citep{skc09,Nandi2012,Deb2013}. 

\citealt{Titarchuk2007} evoked the idea of the diffusive propagation
of perturbation in the disk like configuration of Keplerian (as extended disk) and sub-Keplerian 
(inner part of the disk as `Compton' cloud) matter distribution \citep{ST95}, to understand 
the different noise component in the power spectra. So, it could be possible to explain the evolution of total rms as well as
modeling the broadband power spectra of 
different branches of HID of XTE J1859$+$226, in the context of TCAF 
model, although detailed modeling of evolution of the power 
spectra of the source is beyond the scope of the present work.

The evolution of the HID (Figure \ref{q-dia}) shows that during the rising phase, the 
source was initially in the hard state (region AB),  
and the variation of the spectral parameters, suggest that the spectra is mostly dominated by the 
sub-Keplerian component (i.e., larger `Compton' cloud) with negligible contribution from the Keplerian 
disk. As the source moves towards the hard-intermediate state (implied by Figures \ref{q-dia} and 
\ref{spec-evo_F1234}), we observe that thermal flux increases relatively more than the non-thermal 
flux (Figure \ref{flux-evo_F123}) and both 
are correlated with the increase in frequency of type C QPOs (see Figure \ref{flux-qpo}).
This implies that the Keplerian flow starts increasing along with the sub-Keplerian flow, which is 
still dominating over the Keplerian in the hard-intermediate state (BC of HID). As a result, the energy 
spectra become softer and the size of the CENBOL also starts 
decreasing (see Figure \ref{qpo-pos}). This is evident from the increase of QPO frequencies during 
the hard-intermediate state. It is observed that as the source move from hard to hard-intermediate 
states, the fold energy also (see Figure \ref{fold-ene}) increases (increase in cut-off energy during 
hard to soft transition was observed by \citealt{Fari2013}). The increase in fold energy seems to be 
quite `unnatural' compared to other BH sources, like GX 339$-$4 and XTE J1550$-$564, where Figure 6 
of \citealt{Motta2009} and Figures 1 and 2 of \citealt{TitSha2010} respectively, pointed 
out the monotonic decrease of cutoff energy during the rising phase.
It seems that the shock acceleration mechanism, which converts fraction of thermal electron to the 
non-thermal electron within the Compton cloud, becomes more important as the source 
evolves in this phase, and hence the fold energy increases although the source spectra 
becomes softer and softer. Similar kind of behaviour is observed in the hard and soft state 
spectrum of Cyg X$-$1, which has been modeled using shock acceleration mechanism in 
TCAF \citep{ChakSam2006}.

It has to be noted that the kink observed around $3^{rd}$ day during the rising phase 
(see Figure \ref{qpo-pos}), when the QPO frequency changes from 
3.05 Hz to 3.64 Hz, is peculiar in sense that at the same time there is an abrupt change in the 
photon index from 2.05 to 2.2 along with an increase in disk flux from 
0.42 $\times$ 10$^{-9}$ erg cm$^{-2}$ s$^{-1}$ to 0.7 $\times$ 10$^{-9}$ erg cm$^{-2}$ s$^{-1}$. 
This could be possible due to sudden change in the accretion flow dynamics during this phase.

During the soft-intermediate state (CD \& EF) the evolution pattern of HID is complex. 
We find that the photon index varies in between 2.5 to 2.1 and disk temperature varies
around 0.9 keV, whereas variation of diskbb and power-law flux is not at all correlated 
(see Figure \ref{flux-evo_F123}). 
This unnatural variation of flux occurs in short time-scales (few hrs to day scales). 
Also, the fold energy varies randomly about an average value of 130 keV within 1$\sigma$ error 
limits, whereas the change in fold energy during rising phase is well above the 2$\sigma$ error limit. 
This could be due to multiple ejections that have taken place, and they are 
directly associated with the disk dynamics. We observe all types of QPOs during this phase of which 
the B-type QPO correlates with power-law flux 
whereas its variation is random with disk flux and hence no correlation exists. C-type QPOs are 
correlated with the disk flux but does not show any specific correlation 
with the power-law flux. So, it is possible that sub-Keplerian flow plays a major role 
(to generate various types of QPOs, relation between various types of QPOs with flux) compared to 
the Keplerian flow, as the `hot' Compton cloud evolves faster (i.e., less viscous) than the 
Keplerian disk. 

We could observe a short presence of a `soft state' 
during branch DE for a few observations (see Figure \ref{q-dia}) where the hardness 
ratio is of $\sim$0.1 and total rms of PDS $\sim$ 1\% to 2\% without any signature of QPO. 
During this phase, the Keplerian flow is supposed to dominate over the sub-Keplerian flow to produce 
`softest' spectrum. As observed, the energy spectra is dominated by thermal emission (see 
Figure \ref{flux-evo_ratio}) but harder in nature (disk temperature also less) compared to the 
soft-intermediate state (see Figure \ref{spec-evo_F1234} and \ref{spec-evo_F5-6}). Hence, 
the observed state could not be a `canonical' soft state. This could be possible due to enhancement of 
sub-Keplerian flow during this time that makes the spectra harder along with the strong presence of
Keplerian flow (i.e., thermal emission). It can be understood in the context of TCAF model that, 
the multiple ejections during the soft-intermediate state restrict the disk to become Keplerian 
dominated flow since the ejections are coupled with the Comptonized corona, which is mostly 
sub-Keplerian in nature.

While the source continued in the soft-intermediate state (branch EF),  
there is an indication of weak increase in X-ray intensity (`secondary' emission) of around 24 days 
with peak emission on $69^{th}$ day of the outburst (see Figure \ref{rlc-pca}). 
The monotonic increase in QPO frequencies (see Figure \ref{qpo-evo}) and increase in other spectral 
parameters (Figure \ref{spec-evo_F5-6}, \ref{flux-evo_F123} and Table \ref{secondary}; see also 
\citealt{Dunn2011a}), indicate that there could have been sudden increase in Keplerian as well as 
in sub-Keplerian flow during the declining phase of the outburst, which results as a weak `secondary' 
emission. The weak presence of soft state and 
the weak `secondary' emission, makes the HID different 
from the `standard' evolution pattern of other outbursting sources (\citealt{HB2005}; 
\citealt{TB2005}; \citealt{TB2010}; \citealt{Mandal2010}; 
\citealt{Nandi2012}).

At the end of the outburst, the source transits from soft-intermediate to hard via hard-intermediate 
state before reaching the quiescence phase. The photon index starts decreasing, with a
very weak signature of thermal emission (Figure \ref{spec-evo_F5-6} and \ref{flux-evo_F123}), 
which implies that the emission component is again dominated by sub-Keplerian matter over 
Keplerian matter. 

The modeling of the HID (i.e., q-track, see also \citealt{Mandal2010,Nandi2012}) which follows 
the evolution of X-ray features of the outburst, and of the broadband spectrum (3 - 150 keV) will be 
carried out using TCAF model and will be presented elsewhere (\citealt{Nmo}).

Results of the modeling (see \S 3.2) of QPO evolution based on POS solution and $\Gamma$-QPO 
correlation suggest that the mass of the source is $\sim$ 6.58 to 8.84 M$_{\odot}$. This matches well 
with the previous estimate of the mass as 7.7 $\pm$ 1.3 M$_{\odot}$ by \citealt{Shap2009}.

\begin{figure}
\includegraphics[width=8.5cm]{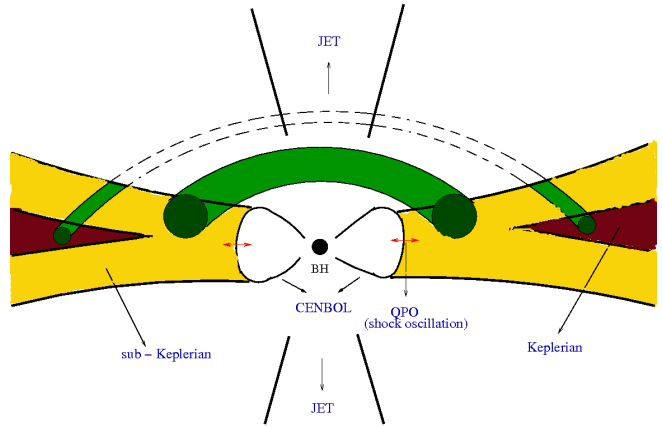}
\caption{A pictorial representation of the magnetised-TCAF model, showing both the Keplerian and 
sub-Keplerian flows along with the toroidal magnetic flux tube and jet emission. (Adopted from 
\citealt{ST95,Nandi2001}). See \citealt{GC2013} for simulation of TCAF model.}
\label{mtcaf}
\end{figure}

The sudden occurrence of a radio flare can be due to some disturbance in the disk system, which  
results in ejection of the material as jets. Several attempts have been made until now, to explain the 
phenomena of jet ejection in BH binaries (see also \citealt{FHB09}). Study of the jet emissions, 
using MHD simulations were carried out by \citealt{Meier2004}, \citealt{MG2004} and \citealt{dV2005}. 
These models suggest that the 
corona is like a windy hot material which blows away as jets from the inner regions of the accretion 
flow. In this work, we attempted to explain the ejection mechanism based on dynamics of magnetic flux 
tube inside the TCAF (see Figure \ref{mtcaf}) as described by \citealt{Nandi2001}. 

When the ejection takes place due to the collapse of magnetic flux tube, the CENBOL gets disrupted and 
the matter is ejected as a jet. As a result, oscillation 
of the CENBOL (i.e., shocked surface) ceases and hence, the QPOs will not be observed 
in the PDS. The energy spectra shall be dominated by thermal emission from the disk, as the 
matter from the CENBOL (`hot' electron) is ejected. The fold energy is observed to 
increase just before the flare and decrease just after the flare. This could be the
possible indication of `disruption' of the corona/CENBOL.
The subsequent observation of QPO and spectral hardening in a time scale of 
few hours to days suggest that the matter flows in sub-Keplerian disk forming the 
Compton cloud/CENBOL responsible for QPO generation. 

As the observations suggest that the companion is of K-type \citep{Corral2011} and hence 
magnetically active, it is possible that the strong magnetic field anchored with the matter is 
continuously fed into the disk system during the rising phase of the outburst. Hence, during the 
transition from hard to soft-intermediate state, the field gets sheared and stronger (in the form 
of toroidal flux tube) and disrupts the Comptonized corona (in different time scales), resulting in 
multiple ejections at different time-scales.

For the flare F1, we observe a type C QPO of 5.97 Hz which corresponds to a CENBOL size of 21.2 r$_g$. 
During the partial absence of the corona (i.e., QPO is not observed in 2 - 5 keV and 13 - 25 keV bands)
 $\sim$ 7 hrs before the flare peak, we observe a 7.79 Hz QPO at a shock location of 17.8 r$_g$. 
Since the sub-Keplerian flow moves in shorter time scale it takes less duration of time to form
 the CENBOL again, which starts oscillating. Hence QPO of 6.1 Hz (type B) is 
observed 2 hrs after the flare peak at a shock location of 20.9 r$_g$. Thus, a duration 
of 9 hrs (maximum) was required for the CENBOL to get disrupted and form again.   

During flare F2, a 7.7 Hz type C* QPO is observed at 17.9 r$_g$ around 21 hrs before the flare peak.
But within 7 hrs, the sub-Keplerian component (as the flow is less viscous) has formed the corona and 
the oscillations of 5.87 Hz (type B) are re-stored at 21.4 r$_g$. A closer look into the PDS evolution 
(Figure \ref{pds-F2}) shows that the QPO frequency increases with diminishing power, before the flare 
occurs and the QPO is not observed (see \S3.3.4).
The QPO was again observed before the radio flare peak, which seems to be inconsistent as compared to 
other flares, where we observe a complete absence of QPO before the radio peak. 
We observe similar characteristics in other sources like XTE J1550$-$564, where the time gap between 
QPO being not observed and the flare peak was more than 24 hrs (\citealt{RNS2013b}).
So, it is possible that the process of evacuation of the CENBOL is rather slow during F2, 
and hence the ejected material took long time to peak as a radio flare. By that time, the corona is 
formed again to reproduce the QPOs, which are interestingly 
of {\it Cathedral type}. As we mentioned, the QPO and Jet formation are coupled with the Comptonized 
corona, which is sub-Keplerian in nature. Hence, the disk-jet coupling mechanism in F2, which seems
to be `complex' in nature (`spectro-temporal' properties are totally different from other flares) could 
be because the dynamics (i.e., the amplification of magnetic field, motion of tube etc.) of flux 
tube within the CENBOL are different.
It requires further investigation to study the coupling between flux tubes and sub-Keplerian flow and 
hence to understand the `complex' disk-jet symbiosis.
But, for both the flares F1 and F2, the disruption of the inner part of disk and
its re-filling time required for oscillation of CENBOL, is around 7 to 9 hrs. So, it is evident that 
within the hours time-scale only the sub-Keplerian 
matter can easily move in, whereas the Keplerian flow does not since it moves in viscous time-scale. 

Before the peak of flare F3 a 7.24 Hz type C* QPO is observed corresponding to the CENBOL size of 
18.6 r$_g$. When the QPO is observed again at 4.87 Hz (type B) the shock location is 24.3 r$_g$. We 
find that, 
the disruption and refilling time of the inner disk (formation of oscillating CENBOL) is $\sim$ 3 days.
It is possible that the magnetic field would have been strong enough (i.e., continuous collapse of
flux tube) to disrupt the disk repeatedly, resulting in multiple signature of radio activity 
(see bottom panel of Figure \ref{rlc-pca}). During 
this long disruption process, no lags in soft to hard photons are observed, disk temperature remains 
almost constant with increased disk flux and the fold energy decreases. These indicate that 
the `hot' Compton cloud is ejected in the form of jets. As the source is also in the declining phase, 
the matter supply in both the form of Keplerian and sub-Keplerian flow has reduced and 
hence, took longer duration to form the inner-part of the disk for the 
oscillation of the CENBOL to start again.

During F4, before the flare peak the QPO was observed at 4.43 Hz (type B) and the CENBOL size is of 
25.9 r$_g$. When the QPO is observed again its frequency is 6.97 Hz (type C*) at 19.1 r$_g$. Before F5, 
we observed a significant QPO of 4.7 Hz (type C*) at 24.9 r$_g$ followed by weak indication of 7.3 Hz 
QPOs (type C*) at 18.2 r$_g$,  and after the flare the QPO is again observed 
with frequency of 5.02 Hz (type C*) at 23.8 r$_g$. The flares F4 and F5 occurred when the source was 
in the declining phase of the outburst, where the 
sub-Keplerian flow dominates. While in decay the rate of incoming matter (both the 
Keplerian and sub-Keplerian) from the companion reduces, and hence the strength of the magnetic field 
also reduces. So, we notice that the radio flares observed have less peak flux as compared to 
previous flares. Although there are no radio observations during the flare F5 
(see \citealt{Brocksopp2002,FHB09}), the results we obtained (\S 3.3.5) suggest that the flare would 
have probably occurred after the day reported by \citealt{Brocksopp2002}. As a result, 
during both the flares, the disruption and refilling time of inner part of the disk (formation of 
oscillating CENBOL) is longer of 
around 30 hrs for F4 (with a possibility of secondary emission after $\sim$ 16.8 hrs with radio 
flux of $\sim$ 8.9 mJy as shown in Figure \ref{rlc-pca}) and 10 days for 
F5 respectively.

We found a possibility of occurrence of another radio flare (i.e., F6) on or before MJD 51499.10 with 
radio flux of $\sim$ 7.9 mJy. There 
is an indication of QPO not being observed after the peak radio flux.  
As there is lack of X-ray and radio observations during this time, it is not
possible to pinpoint the exact time of flare ejection.

Above discussions on all the flares suggest that the Radio flare take different time to 
peak and QPOs are not observed during the flares. The re-filling time scale ($\sim$ few hrs to days 
based on available data) to form oscillating CENBOL (corona) resulting in QPOs, also suggests that 
the accretion flow could be sub-Keplerian in nature.

Similar study of the disk-jet connection based on the magnetized TCAF model has been earlier presented 
for GRS 1915$+$105 by \citealt{Nandi2001, SV2001}.

We observe that the type of QPO is independent of the ejection event (See \S 3.3 and Column 10 
of Table \ref{tabflares}). The results mentioned in \S 3.3 clearly show that the QPOs observed before 
and after the flare, need not be of specific type. But, the prediction of Jet ejections (without 
any radio detection), based on QPOs being observed or not in the power spectra, requires further 
studies in the context of TCAF model which could be possible after detailed investigation of X-ray 
observation of other outbursting sources. 

We have also observed similar characteristics (but with no definite relation between the types of QPOs 
before and after the ejection) in other BH sources. In the case of XTE J1550$-$564, GRO J1655$-$40 and 
XTE J1752$-$223, we found that the QPO observed before and after the flare are of type C. For 
H 1743$-$322, we observed a type C QPO before the ejection and the PDS showed a type B QPO after the 
ejection. So, in the present scenario, it is not possible to generalise or establish a relation 
between types of QPO and flare ejection events. Although, there were attempts to establish the link of 
B type QPOs with radio flare, along with the drop in rms in the power spectra 
(see \citealt{Sol08,FHB09}). 
Detailed study on QPO types and radio flares will be performed for several BH sources (based on 
available X-ray and radio data) (\citealt{RNS2013b}).

Due to the complex behaviour of QPOs during ejection events, we presented in \S 3.4 how
different types of QPOs are correlated with the flux variations. 
The variation of type B/C QPOs w.r.t disk and powerlaw flux is not similar to that observed for 
GX 339$-$4 by \citealt{Motta2011}, where both type B and C QPOs are correlated with the disk and 
powerlaw flux (see Figures 5 and 8 in \citealt{Motta2011}). Figure \ref{flux-qpo} 
also shows that the type C* QPOs have a weak correlation with the disk flux but does not have any 
correlation with the powerlaw flux, and a 
similar characteristic was observed for GX 339$-$4 by \citealt{Motta2011}. Type A QPOs do not show 
any correlation with flux for GX 339$-$4, while XTE J1859$+$226 has only very few type A QPOs, which 
limits the results. \citealt{Sob2000} and \citealt{M09} observed a positive correlation between QPO 
frequency and disk flux during the rising phase (for GRO J1655$-$40, XTE J1550$-$564, H 1743$-$322), 
which is also observed for XTE J1859$+$226. An opposite correlation was 
observed between QPO frequency and powerlaw flux for GRO J1655$-$40, while a positive 
correlation was observed for XTE J1550$-$564 and H 1743$-$322.

In the present work, we have not explained the origin of different types of QPOs before and after the 
ejections. According to TCAF model, the base of the Jet and origin of QPOs both are directly 
coupled with the CENBOL dynamics. Recently, we have done simulation to show that the mass loss from 
the disk and origin of QPOs in BH are linked with the CENBOL (\citealt{dcnm14}, see references therein).
So, it could be possible to address the types of QPOs during the ejections, as the QPO generation in 
TCAF model depends on the sub-Keplerian flow characteristics (i.e., viscosity, angular momentum, 
energy etc. of the flow). This work which involves detailed simulation (SPH and TVD based) is under
progress and would be presented elsewhere.

There have been reports of flare events (radio ejections) during the outburst of sources like 
XTE J1748$-$288 (\citealt{Brocksopp2007}), XTE J1752$-$223 (\citealt{Brock2010}) 
and H 1743$-$322 (\citealt{h2012}). 
The study of the outburst evolution and flaring events associated with 
X-ray properties observed for the BH sources XTE J1748$-$288, XTE J1752$-$223, H 1743$-$322, 
XTE J1550$-$564, GRO J1655$-$40, GX 339$-$4 and MAXI J1836$-$194, in the context of TCAF model is 
under progress. 
Preliminary results on the presence and absence of QPOs associated 
with flaring events for several outbursting black holes have been presented in \citealt{RNS2013,NRS2013}.

\section{Summary}

Based on the findings of `spectro-temporal' evolution of the source XTE J1859$+$226, we can 
summarize that: 

\begin{itemize}

\item The source evolved through various states in HID with a sequence of {\it hard $\rightarrow$ 
hard-intermediate $\rightarrow$ soft-intermediate $\rightarrow$ soft state 
(not `canonical') $\rightarrow$ soft-intermediate $\rightarrow$ hard-intermediate 
$\rightarrow$ hard states} before 
reaching to the quiescence phase. Multiple radio ejections (or the enhancement in sub-Keplerian flow) 
might have stopped the source to evolve from soft-intermediate to `canonical' soft state.

\item The QPO frequency increases monotonically during the rising phase of the outburst. The 
evolution is modeled with POS solution to estimate the size of the `Compton' corona. Similar kind 
of variation is also observed in other outbursting BH sources.

\item Results from the modeling of the QPO evolution and $\Gamma$-QPO correlation suggest that mass 
of the source could be between 6.58 M$_{\odot}$ and 8.84 M$_{\odot}$.

\item The temporal evolution of the PDS shows that the QPOs are not observed, during the 
ejection as observed in radio.

\item During F1, no QPO is observed in the soft band of 2 - 5 keV (also not in 13 - 25 keV band), 
but a broad QPO of less amplitude is observed in the 6 - 13 keV band. This implies that there has 
been a `partial' absence of the QPO in 2 - 5 keV and 13 - 25 keV bands.

\item During F2, F3, F4 and F5, we find that the QPOs are not at all observed in the PDS.

\item The QPO frequency observed before and after the flare need not be of same type during all 
the flares.

\item During all the ejections, the X-ray spectrum gets soften and fold energy ($E_{fold}$) decreases, 
as the `hot' electron cloud (corona) gets disrupted and spectrum is dominated by disk emission. 
Only in the case of F2, $E_{fold}$ increases, the evacuation process seems to be complex.

\item `Disruption' of inner part of the disk (corona/CENBOL) occurs as a resultant of dominant thermal 
emission over non-thermal flux resulting in QPOs being not observed. This could be due to the 
`catastrophic collapse' of toroidal flux tube of strong magnetic field in the hot region of 
sub-Keplerian flow.

\item During the declining phase, the source might have undergone a `secondary' outburst 
observed as a weak X-ray flaring activity for $\sim$ 24 days.

\item Due to lack of continuous X-ray observation during the flare time, it was not possible to 
tighten the duration of disruption and refilling time scale. So, it is required to have more
precise and simultaneous observation in {\it radio-UV-X-rays} to provide better picture of 
disk-jet dynamics in outbursting sources. In this context, continuous monitoring of outbursting 
sources with India's upcoming multi-wavelength satellite ASTROSAT along with GMRT, VLBI etc. 
will be a best opportunity for studying the complex accretion dynamics 
of BH sources.

\end{itemize} 


\section*{Acknowledgments}

We thank Dr. P. Sreekumar  (IIA, Bangalore), Dr. S. Seetha (ISRO Headquarters) and Dr. A. Agarwal (ISAC, Bangalore) for various suggestions and support. Radhika D., acknowledges the research fellowship provided by ISRO Satellite Centre.

This research has made use of the data obtained through High Energy Astrophysics Science Archive 
Research Center on-line service, provided by NASA/Goddard Space Flight Center and of the General 
High-energy Aperiodic Timing Software (GHATS) package developed by Dr. Tomaso Belloni at 
INAF - Osservatorio Astronomico di Brera. 

We thank Dr. Dipankar Bhattacharya of Inter University Centre for Astronomy \& Astrophysics, Pune for 
providing an opportunity to participate in the `Advanced X-ray timing workshop' and also to Dr. Tomaso 
Belloni for the sessions on usage of GHATS. 

We are thankful to the anonymous referees for their suggestions which helped to improve the manuscript.



\section*{References}
\begin{thebibliography}{}

\bibitem[\protect\citeauthoryear{Belloni \& Hasinger}{1990}]{BH1990}Belloni T. M., Hasinger G., 1990, A\&A, 230, 103
\bibitem[\protect\citeauthoryear{Belloni et al.}{2002}]{Belloni2002}Belloni T. M., Psaltis D., van der Klis M., 2002, ApJ, 572, 392
\bibitem[\protect\citeauthoryear{Belloni et al.}{2005}]{TB2005}Belloni T. M., Homan J., Casella P., et al., 2005, A\&A, 440, 207
\bibitem[\protect\citeauthoryear{Belloni et al.}{2006}]{TB2006}Belloni T. M., Parolin I., Del Santo M., et al., 2006, MNRAS, 367, 1113
\bibitem[\protect\citeauthoryear{Belloni }{2010}]{TB2010}Belloni T. M., 2010, `The Jet Paradigm - From Microquasars to Quasars', Lect. Notes Phys 794, edited by T. Belloni, 794, 53
\bibitem[\protect\citeauthoryear{Blandford \& Znajek}{1977}]{BZ1997}Blandford R. D., Znajek R. L., 1977, MNRAS, 179, 433
\bibitem[\protect\citeauthoryear{Bradt et al.}{1993}]{BRS93}Bradt H. V., Rothschild R. E., Swank J. H., 1993, A\&A supplement series, 97, 355
\bibitem[\protect\citeauthoryear{Brocksopp et al.}{2002}]{Brocksopp2002}Brocksopp C., Fender R. P., McCollough M., et al., 2002, MNRAS, 331, 765
\bibitem[\protect\citeauthoryear{Brocksopp et al.}{2007}]{Brocksopp2007}Brocksopp C., Miller-Jones J. C. A., Fender R. P., Stappers B. W., 2007, MNRAS, 378, 1111
\bibitem[\protect\citeauthoryear{Brocksopp et al.}{2010}]{Brock2010}Brocksopp C., Corbel, S., Tzioumis T., Fender R. et al., 2010, Atel, 2400 
\bibitem[\protect\citeauthoryear{Cadolle Bel et al.}{2010}]{CB2011}Cadolle Bel M., Rodriguez J., D'Avanzo P. et al., 2011, A\&A, 534, 119
\bibitem[\protect\citeauthoryear{Casella et al.}{2004}]{Casella2004}Casella P., Belloni T., Homan J., Stella L., 2004, A\&A, 426, 587
\bibitem[\protect\citeauthoryear{Casella et al.}{2005}]{Cas2005}Casella P., Belloni T., Homan J., Stella L., 2005, ApJ, 629, 403
\bibitem[\protect\citeauthoryear{Chakrabarti \& D'Silva}{1994}]{CD94}Chakrabarti S. K., \& D'Silva, S., 1994, ApJ, 424, 138
\bibitem[\protect\citeauthoryear{Chakrabarti \& Titarchuk}{1995}]{ST95}Chakrabarti S.K., Titarchuk L.G., 1995, ApJ, 455, 623
\bibitem[\protect\citeauthoryear{Chakrabarti \& Manickam}{2000}]{CM2000}Chakrabarti S.K. \& Manickam, S.G., 2000, ApJ, 531, L41
\bibitem [\protect\citeauthoryear{Chakrabarti et al.}{2002}]{Chak2002}Chakrabarti, S. K., Nandi, A., Manickam, S., et al., 2002, ApJ, 579, L21
\bibitem[\protect\citeauthoryear{Chakrabarti et al.}{2004}]{skc04}Chakrabarti S. K., Acharya K., Molteni D., 2004, A\&A, 421, 1
\bibitem[\protect\citeauthoryear{Chakrabarti \& Mandal}{2006}]{ChakSam2006}Chakrabarti S. K., Mandal S., 2006, ApJ, 642, 49
\bibitem[\protect\citeauthoryear{Chakrabarti et al.}{2008}]{skc08}Chakrabarti S. K., Debnath D., Nandi A. \& Pal P. S., 2008, A\&A, 489, L41
\bibitem[\protect\citeauthoryear{Chakrabarti et al.}{2009}]{skc09}Chakrabarti S. K., Dutta B. G., \& Pal P. S., 2009, MNRAS, 394, 1463
\bibitem[\protect\citeauthoryear{Corral-Santana et al.}{2011}]{Corral2011}Corral-Santana J. M., Casares J., Shahbaz T., et al., 2011, MNRAS, 413, L15
\bibitem[Das et al.(2014)]{dcnm14}Das S., Chattopadhyay I., Nandi A., Molteni D., 2014, MNRAS, 442, 251
\bibitem[\protect\citeauthoryear{De Villiers et al.}{2005}]{dV2005}De Villiers J. P., Hawley J. F., Krolik J. H., Hirose S., 2005, ApJ, 620, 878
\bibitem[\protect\citeauthoryear{Debnath et al.}{2008}]{Deb2008}Debnath D., Chakrabarti S. K., Nandi A., 2008, BASI, 36, 151 
\bibitem[\protect\citeauthoryear{Debnath et al.}{2013}]{Deb2013}Debnath D., Chakrabarti S. K., Nandi A., 2013, AdSpR, 52, 2143
\bibitem[\protect\citeauthoryear{Dunn et al.}{2011a}]{Dunn2011a}Dunn R. J. H., Fender R. P., Kording E. G., et al., 2011, MNRAS, 403, 61
\bibitem[\protect\citeauthoryear{Dunn et al.}{2011b}]{Dunn2011b}Dunn R. J. H., Fender R. P., Kording E. G., et al., 2011, MNRAS, 411, 337
\bibitem[\protect\citeauthoryear{Ebisawa et al.}{1994}]{Ebisawa1994}Ebisawa K., Ogawa M., Aoki T., et al., 1994, PASJ, 46, 375
\bibitem[\protect\citeauthoryear{Esin et al.}{1997}]{Esin1997}Esin A. A., McClintock J. E., Narayan R., 1997, ApJ, 489, 865
\bibitem[\protect\citeauthoryear{Farinelli et al.}{2013}]{Fari2013}Farinelli R., Amati L., Shaposhnikov N., et al., 2013, MNRAS, 428, 3295
\bibitem[\protect\citeauthoryear{Fender et al.}{2004}]{FBG04}Fender R. P., Belloni T., Gallo E., 2004, MNRAS, 355, 1105
\bibitem[\protect\citeauthoryear{Fender et al.}{2009}]{FHB09}Fender R. P., Homan J., Belloni T., 2009, MNRAS, 396, 1307
\bibitem[\protect\citeauthoryear{Feroci et al.}{1999}]{Feroci99}Feroci M., Matt G., Pooley G., Costa E., et al., 1999, A\&A, 351, 985
\bibitem[\protect\citeauthoryear{Garnavich et al.}{1999}]{Garn99}Garnavich P. M., Stanek K. Z., Berlind P., 1999, IAUC, 7276
\bibitem[\protect\citeauthoryear{Gierlinski \& Done}{2004}]{GD2004}Gierlinski M., Done C., 2004, MNRAS, 347, 885
\bibitem[\protect\citeauthoryear{Giri \& Chakrabarti}{2013}]{GC2013}Giri K., Chakrabarti S. K., 2013, MNRAS, 430, 2836
\bibitem[\protect\citeauthoryear{Homan \& Belloni}{2005}]{HB2005}Homan J., Belloni T., 2005 Ap\&SS, 300, 107
\bibitem[\protect\citeauthoryear{Ingram \& Done}{2011}]{ID2011}Ingram A., Done C., 2011, MNRAS, 415, 2323 
\bibitem[\protect\citeauthoryear{Lee et al.}{2011}]{LRC2011}Lee, S., Ryu, D., Chattopadhyay, I., 2011, ApJ, 728, 142
\bibitem[\protect\citeauthoryear{Mandal \& Chakrabarti }{2010}]{Mandal2010}Mandal S., Chakrabarti S. K., 2010, ApJ, 710, L147
\bibitem[\protect\citeauthoryear{Markwardt}{2001}]{Mark2001}Markwardt C. B., 2001, Astrophysics \& Space science, 276, 209
\bibitem[\protect\citeauthoryear{Markwardt}{2009}]{Markmp}Markwardt C. B., 2009, ASPC, 411, 251
\bibitem[\protect\citeauthoryear{McClintock \& Remilard}{2006}]{RM04}McClintock J. E., Remilard R. A., 2006, `Black hole binaries', Compact Stellar X-ray sources, edited by Lewin W. H. G. and M. van der Klis 
\bibitem[\protect\citeauthoryear{McClintock et al.}{2009}]{M09}McClintock J. E., Remilard R. A., Rupen M. P., et al., 2009, ApJ, 698, 1398
\bibitem[\protect\citeauthoryear{McCollough \& Wilson}{1999}]{McW99}McCollough M. L., Wilson C. A., 1999, IAUC, 7282
\bibitem[\protect\citeauthoryear{McKinney \& Gammie et al.}{2004}]{MG2004}McKinney J. C., Gammie C. F., 2004, ApJ, 611, 977
\bibitem[\protect\citeauthoryear{Meier \& Nakamura}{2004}]{Meier2004}Meier D. L., Nakamura M., Proceedings 3-D signatures in Stellar explosions, ed. P. Hoeflich, P. Kumar, C. Wheeler, 2004, Cambridge University press, 219
\bibitem[\protect\citeauthoryear{Merloni et al.}{2000}]{Mer2000}Merloni A., Fabian A. C. and Ross R. R., 2000, MNRAS, 313, 193
\bibitem[\protect\citeauthoryear{Meyer et al.}{2007}]{Meyer2007}Meyer F., Liu B. F., Meyer-Hofmeister E., 2007, A\&A, 463, 1
\bibitem[\protect\citeauthoryear{Meyer-Hofmeister et al.}{2009}]{Meyer2009}Meyer-Hofmeister E., Liu B . F, Meyer F., 2009, A\&A, 508, 329
\bibitem[\protect\citeauthoryear{Miller-Jones et al.}{2012}]{h2012}Miller-Jones J. C. A., Sivakoff G. R., Altamirano D., et al., 2012, MNRAS, 421, 468
\bibitem[\protect\citeauthoryear{Miyamoto et al.}{1991}]{Miyamoto91}Miyamoto S., Kimura K., Kitamoto S., et al., 1991, ApJ, 383, 784
\bibitem[\protect\citeauthoryear{Molteni et al.}{1996}]{MSC96}Molteni D., Sponholz H., Chakrabarti S. K., 1996, ApJ, 457, 805
\bibitem[\protect\citeauthoryear{Motta et al.}{2009}]{Motta2009}Motta S., Belloni T., Homan J., 2009, MNRAS, 400, 1603
\bibitem[\protect\citeauthoryear{Motta et al.}{2011}]{Motta2011}Motta S., Munoz-Darias T., Casella P., et al., 2011, MNRAS, 418, 2292
\bibitem[\protect\citeauthoryear{Munoz-Darias et al.}{2010}]{MD2010}Munoz-Darias T., Motta S. and Belloni T., 2010, MNRAS, 410, 679
\bibitem[\protect\citeauthoryear{Nandi et al.}{2001}]{Nandi2001}Nandi A., Chakrabarti S. K., Vadawale S. V., Rao A. R., 2001, A\&A, 380, 245
\bibitem[\protect\citeauthoryear{Nandi et al.}{2012}]{Nandi2012}Nandi A., Debnath D., Mandal S., Chakrabarti S. K., 2012, A\&A, 542, 56
\bibitem[\protect\citeauthoryear{Nandi \& Radhika}{2012}]{NR2012}Nandi, A. \& Radhika. D, 2012, COSPAR Scientific Assembly, Mysore, INDIA
\bibitem[\protect\citeauthoryear{Nandi et al.}{2013}]{NRS2013}Nandi, A., Radhika. D, Seetha. S., 2013, ASI Conference Series, 8, 71 (arxiv : 1308.4567)
\bibitem[\protect\citeauthoryear{Nandi et al. (in prep.)}{}]{Nmo} Nandi A. et al. (in prep.)
\bibitem[\protect\citeauthoryear{Okuda et al.}{2007}]{Okuda2007}Okuda T., Teresi V., Molteni D., 2007, MNRAS, 377, 1431
\bibitem[\protect\citeauthoryear{Pooley \& Hjellming}{1999}]{Pooley99}Pooley G. G., Hjellming R. M., 1999, IAUC, 7278
\bibitem[\protect\citeauthoryear{Radhika et al.}{2013}]{RNS2013}Radhika. D, Nandi A., S. Seetha, 2013, MmSAI, 84, 624 (arxiv : 1301.7234)
\bibitem[\protect\citeauthoryear{Radhika et al. (in prep.)}{}]{RNS2013b}Radhika. D et al. (in prep.)
\bibitem[\protect\citeauthoryear{Remilard et al.}{1999}]{Rem1999}Remilard R. A. et al., 1999, ApJ, 552, 397
\bibitem[\protect\citeauthoryear{Rodriguez \& Prat}{2008}]{Rodriguez2008}Rodriguez J., Prat L., 2008, Proceedings of Science, 7th INTEGRAL workshop (arxiv : 0811:3519)
\bibitem[\protect\citeauthoryear{Rodriguez \& Varniere}{2011}]{Rodriguez2011}Rodriguez J., Varniere P., 2011, ApJ, 735, 79
\bibitem[\protect\citeauthoryear{Ryu et al.}{1997}]{Ryu97}Ryu D., Chakrabarti S. K., Molteni D., 1997, ApJ, 474, 378
\bibitem[\protect\citeauthoryear{Shakura \& Sunyaev}{1973}]{SS73}Shakura N. I., Sunyaev R. A., 1973, A\&A, 24, 337
\bibitem[\protect\citeauthoryear{Shaposhnikov \& Titarchuk}{2007}]{Shap2007}Shaposhnikov N, Titarchuk L., 2007, ApJ, 663, 445
\bibitem[\protect\citeauthoryear{Shaposhnikov \& Titarchuk}{2009}]{Shap2009}Shaposhnikov N, Titarchuk L.,2009, ApJ, 699, 453
\bibitem[\protect\citeauthoryear{Sobczak et al.}{2000}]{Sob2000}Sobczak G. J., McClintock J. E., Remilard R. A., et al., 2000, ApJ, 531, 537
\bibitem[\protect\citeauthoryear{Soleri et al.}{2008}]{Sol08}Soleri P., Belloni T., Casella P., 2008, MNRAS, 383, 10989
\bibitem[\protect\citeauthoryear{Stiele et al.}{2013}]{Stiele2013}Stiele H., Belloni T. M., Kalemci E., Motta S., 2013, MNRAS, 429, 2655 
\bibitem[\protect\citeauthoryear{Tagger \& Pellat}{1999}]{TagPel} Tagger M., Pellat R., 1999, A\&A, 349, 1003
\bibitem[\protect\citeauthoryear{Titarchuk}{1994}]{Titarchuk1994}Titarchuk L. G., 1994, ApJ, 434, 570
\bibitem[\protect\citeauthoryear{Titarchuk \& Fiorito }{2004}]{TF04}Titarchuk, L. G., Fiorito, R. 2004, ApJ, 612, 988
\bibitem[\protect\citeauthoryear{Titarchuk et al.}{2007}]{Titarchuk2007}Titarchuk L., Shaposhnikov N., \& Arefiev V., 2007, ApJ, 660, 556
\bibitem[\protect\citeauthoryear{Titarchuk \& Shaposhnikov}{2010}]{TitSha2010}Titarchuk L., Shaposhnikov N., 2010, ApJ, 724, 1147
\bibitem[\protect\citeauthoryear{Titarchuk \& Osherovich}{2000}]{TitOsh2000}Titarchuk L., Osherovich V., 2000, ApJ, 542, L111
\bibitem[\protect\citeauthoryear{Vadawale et al.}{2001}]{SV2001}Vadawale S. V., Rao A. R., Nandi A., Chakrabarti S. K., 2001, A\&A, 370, L17
\bibitem[\protect\citeauthoryear{Wagner et al.}{1999}]{Wagner99}Wagner R. M., Schmidt G. D., Shrader C. R., 1999, IAUC, 7279
\bibitem[\protect\citeauthoryear{Wood et al.}{1999}]{Wood99}Wood A., Smith D. A., Marshall F. E., Swank J. H., 1999, IAUC, 7274
\bibitem[\protect\citeauthoryear{Yang et al.}{2010}]{Yang2010}Yang J., Brocksopp C., Corbel S., et al., 2010, MNRAS, 409, L64
\bibitem[\protect\citeauthoryear{Zhang et al.}{1995}]{Zhang1995}Zhang W., Jahoda K., Swank J. H., et al., 1995, ApJ, 449, 930

\end{thebibliography}
\end{document}